\RequirePackage{lineno}
\documentclass[twocolumn,showpacs,superscriptaddress,amsmath,amssymb,nofootinbib]{revtex4-2}
\usepackage[x11names]{xcolor}
\usepackage{caption}
\usepackage[colorlinks, linkcolor=blue, citecolor=OliveGreen, urlcolor=magenta]{hyperref}
\usepackage{orcidlink}
\usepackage{ulem}
\allowdisplaybreaks[4]
\usepackage{graphicx}
\usepackage{subfigure}
\usepackage[utf8]{inputenc}
\usepackage{subcaption}
\usepackage{epsfig}
\usepackage{overpic}
\usepackage{dcolumn}
\usepackage{bm}
\usepackage{float}
\usepackage{color}
\usepackage{xspace}
\usepackage{multirow}
\usepackage{epstopdf}
\usepackage{soul}
\usepackage{verbatim}
\usepackage{enumitem}
\usepackage{todonotes}
\usepackage{slashed}
\usepackage{cancel}
\usepackage{array}
\usepackage{amsmath}
\usepackage{booktabs}
\newcommand{\iu}{\bm{\mathrm{i}}}
\newcommand{\gn}{\bm{\mathfrak{n}}}
\newcommand{\gm}{\bm{\mathfrak{m}}}
\newcommand{\gp}{\bm{\mathfrak{p}}}

\usepackage[toc,page]{appendix}

\definecolor{OliveGreen}{rgb}{0.4, 0.8, 0.1}

\usepackage{diagbox}

\newcommand{\moe}{\affiliation{Key Laboratory of Atomic and Subatomic Structure and
Quantum Control (MOE), Guangdong-Hong Kong Joint Laboratory of Quantum Matter, Guangzhou 510006, China}}

\newcommand{\iqm}{\affiliation{State Key Laboratory of Nuclear Physics and
Technology, Institute of Quantum Matter, South China Normal
University, Guangzhou 510006, China}}

\newcommand{\gbrce}{\affiliation{Guangdong Basic Research Center of Excellence for
Structure and Fundamental Interactions of Matter, Guangdong Provincial Key Laboratory of Nuclear Science, Guangzhou 510006, China}}

\newcommand{\scnt}{\affiliation{Southern Center for Nuclear-Science Theory (SCNT), Institute of Modern Physics, Chinese Academy of Sciences, Huizhou 516000, Guangdong Province, China}}

\begin{document}
\title{\boldmath Does $\psi(4660)$ exist? }

\author {\mbox{Xian-Chen Wang}\orcidlink{0009-0004-9492-6767}}
\iqm
\moe
\gbrce

\author {\mbox{Quanxing Ye}\orcidlink{0009-0006-8322-3163}}
\email{Co-first author}
\iqm
\moe
\gbrce

\author {\mbox{Qian Wang\orcidlink{0000-0002-2447-111X}}}
\email{qianwang@m.scnu.edu.cn, corresponding author}
\iqm
\gbrce
\scnt

\date{\today}
\begin{abstract}
The nature of the vector charmoniumlike states $\psi(4360)$ and $\psi(4660)$ has remained controversial for nearly two decades, and the very existence of the $\psi(4660)$ has recently been called into question: the high-precision BESIII measurement of the $e^+e^-\to \Lambda_c^+\bar{\Lambda}_c^-$ cross section reveals a nearly flat line shape around 4.66 GeV with a nonzero value right at threshold, in striking contrast to the near-threshold enhancement reported earlier by Belle. In this work, we address this puzzle within a comprehensive S-wave coupled-channel framework for $e^+e^-$ annihilation in the energy region $\sqrt{s}\in [4.0,5.0]~\mathrm{GeV}$. The formalism incorporates the open-charm baryon-antibaryon channels $\Lambda_c^+\bar{\Lambda}_c^-$, $\Xi_c^+\bar{\Xi}_c^-$, and $\Xi_c^0\bar{\Xi}_c^0$ (organized via SU(3) flavor symmetry), the inelastic $\psi(2S)\pi^+\pi^-$ channel mediated by the $\sigma(500)$ and $f_0(980)$ intermediate resonances, and the bare vector charmonium states $\psi(4360)$ and $\psi(4660)$, assigned as the 4S and 5S excitations. Short-range contact potentials constrained by heavy-quark spin symmetry are employed, the scattering amplitudes are obtained by solving the Lippmann-Schwinger equation. A combined fit to the Belle and BESIII data on $e^+e^-\to \Lambda_c^+\bar{\Lambda}_c^-$ and $e^+e^-\to \psi(2S)\pi^+\pi^-$ yields two poles, at $(4401.93 \pm 16.96\iu) ~\mathrm{MeV}$ and $(4631.39\pm 62.96\iu)~\mathrm{MeV}$, corresponding to the $\psi(4360)$ and $\psi(4660)$, respectively. By tracing the pole trajectories as the coupling to the bare charmonium states is gradually switched off, we demonstrate that both poles originate from the bare $4S$ and $5S$ vector charmonium states strongly renormalized by coupled-channel dynamics. In particular, the $\psi(4660)$ emerges as a broad resonance located about 58~MeV above the $\Lambda_c^+\bar{\Lambda}_c^-$ threshold—its flat experimental line shape is a direct consequence of this threshold interplay, and provides evidence for its existence. Our conclusion is robust against the variation of the regularization cutoff.
\end{abstract}

\maketitle

\section{INTRODUCTION}

Since the discovery of the $X(3872)$ by the Belle Collaboration in 2003~\cite{Belle:2003nnu}, a wealth of exotic hadron candidates---collectively known as XYZ states---have been observed in experiment.
A striking feature of many XYZ states is their proximity to open-flavor thresholds, which highlights the crucial role of nearby thresholds and coupled-channel effects in their formation~\cite{Guo:2017jvc}.
Numerous theoretical interpretations have been advanced, including modified quark models~\cite{Barnes:2005pb,Sultan:2014oua},
threshold cusp effects from nearby thresholds~\cite{Szczepaniak:2015eza},
hadronic molecules~\cite{Swanson:2003tb,Guo:2017jvc,Xiao:2016wbs},
compact tetraquark states~\cite{Maiani:2004vq},
and conventional charmonium assignments~\cite{Achasov:2015oia}. For recent comprehensive reviews of both the experimental and theoretical status of the XYZ states, we refer the reader to Refs.~\cite{Hosaka:2016pey,Brambilla:2019esw,Richard:2016eis,Guo:2019twa,Yuan:2021wpg,Chen:2022asf,Meng:2022ozq,Liu:2024uxn,Esposito:2025hlp,Esposito:2016noz}.

Among these intriguing candidates, the vector state $\psi(4660)$ stands out, as experimental and theoretical investigations have so far produced conflicting findings.
In experiment, the $\psi(4660)$ was first observed by the Belle Collaboration in 2007~\cite{Belle:2007umv} in the $e^+e^- \to \psi(2S)\pi^+\pi^-$ process. In the following year, another enhancement, denoted as $X(4630)$, was reported by Belle in the $e^+e^- \to \Lambda_c^+\bar{\Lambda}_c^-$ cross section~\cite{Belle:2008xmh}. The proximity of their masses and widths suggests that the two structures may correspond to the same state. Subsequent measurements by the BaBar~\cite{BaBar:2012hpr} and Belle~\cite{Belle:2014wyt} Collaborations in $e^+e^-\to\psi(2S)\pi^+\pi^-$ further confirmed clear enhancements consistent with the $\psi(4660)$ in the cross sections. More recently, the BESIII Collaboration has provided precise cross-section measurements at selected center-of-mass energies for the
$e^+e^-\to\psi(2S)\pi^+\pi^-$~\cite{BESIII:2017tqk,BESIII:2021njb} and
$e^+e^-\to\Lambda_c^+\bar{\Lambda}_c^-$~\cite{BESIII:2017kqg,BESIII:2023rwv} processes.
In the $\psi(2S)\pi^+\pi^-$ channel, BESIII observes peak structures near $4230~\mathrm{MeV}$, $4360~\mathrm{MeV}$, and $4660~\mathrm{MeV}$, consistent with those previously reported by Belle~\cite{Belle:2007umv,Belle:2014wyt} and BaBar~\cite{BaBar:2012hpr}; in particular, the mass of the $\psi(4660)$ extracted by BESIII is in good agreement with the earlier results~\cite{BaBar:2012hpr,Belle:2014wyt} in the same process.
Furthermore, evidence for the $\psi(4660)$ resonance has been reported by both Belle~\cite{Belle:2019qoi,Belle:2020wtd} and BESIII~\cite{BESIII:2022yga,BESIII:2023cmv} in various other decay channels; a systematic discussion can be found in Ref.~\cite{Wang:2025dur}.

\begin{figure}[tb]
\centering
\includegraphics[width=0.9\columnwidth]{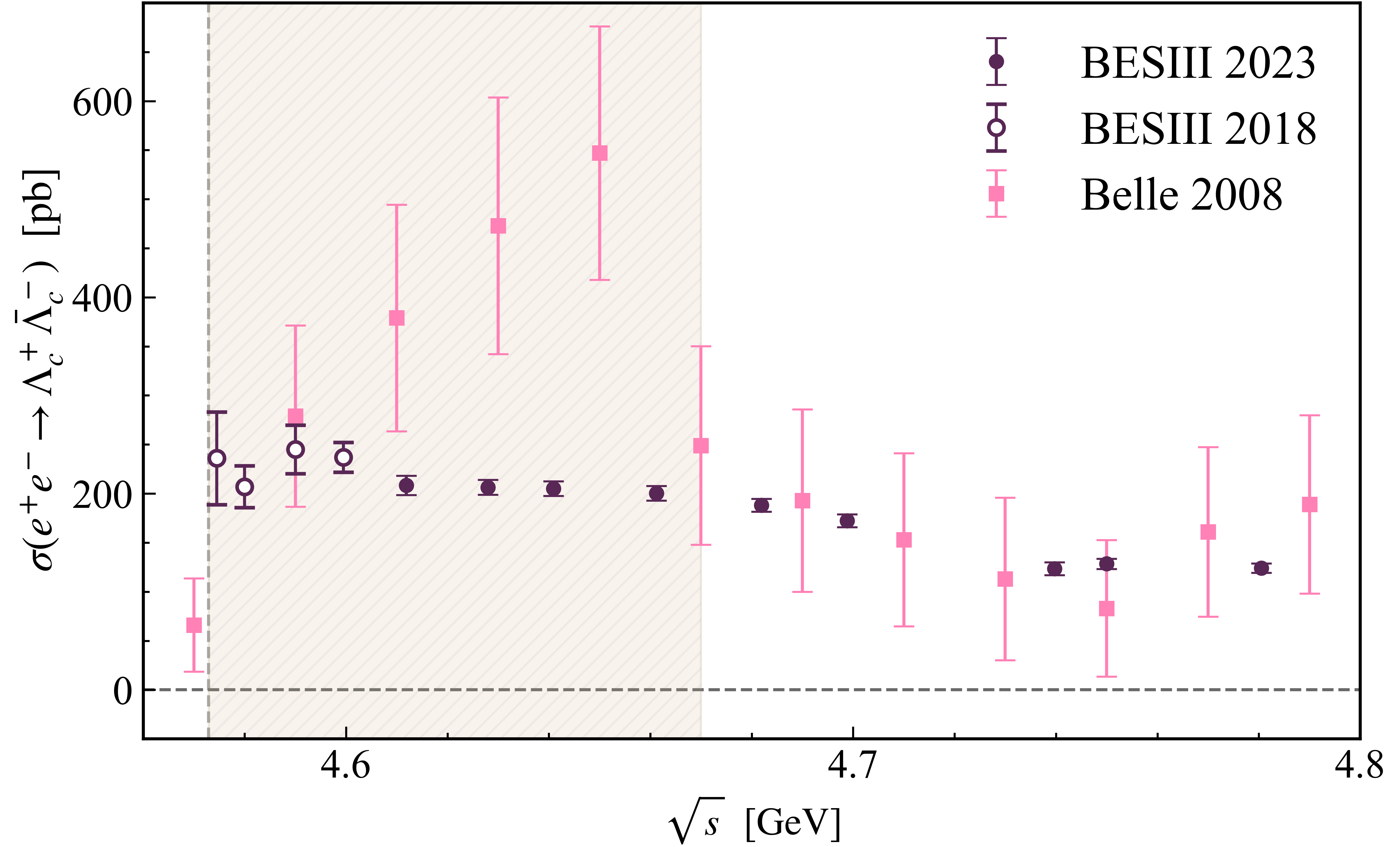}
\caption{Comparison of the $e^+e^-\to\Lambda_c^+\bar{\Lambda}_c^-$ cross section data from Belle~\cite{Belle:2008xmh} and BESIII~\cite{BESIII:2017kqg,BESIII:2023rwv}. The shaded band highlights the region of significant tension between the two datasets above the $\Lambda_c^+\bar{\Lambda}_c^-$ threshold, marked by the vertical dashed line.}
\label{fig:CompLamLam}
\end{figure}

In striking contrast, the BESIII measurements~\cite{BESIII:2017kqg,BESIII:2023rwv} of the $e^+e^-\to\Lambda_c^+\bar{\Lambda}_c^-$ cross section show no evident resonant enhancement and remain nearly flat, in marked tension with the prominent peak structure around $4.66~\mathrm{GeV}$ observed by Belle~\cite{Belle:2008xmh}, as illustrated in Fig.~\ref{fig:CompLamLam}.
In addition, the BESIII measurement exhibits an anomalous enhancement right at the $\Lambda_c^+\bar{\Lambda}_c^-$ threshold, which has been attributed to a Coulomb-like effect arising from strong final-state interactions~\cite{Amoroso:2021slc}, or to the presence of a virtual state~\cite{Cao:2019wwt} or of a bound state below the $\Lambda_c^+\bar{\Lambda}_c^-$ threshold~\cite{Salnikov:2023qnn}. Alternatively, the nontrivial near-threshold behavior could be produced by a mixture of $S$- and $D$-wave components of the $\Lambda_c^+\bar{\Lambda}_c^-$ wave function induced by the tensor interaction~\cite{Milstein:2022bfg}.

On the theoretical side, various interpretations have been proposed to understand the nature of the $\psi(4660)$. These interpretations fall into two categories: a conventional charmonium state~\cite{Badalian:2008dv,Segovia:2008ta,Zhao:2023hxc} or a dynamically generated state~\cite{Cotugno:2009ys,Liu:2016sip,Nakamura:2023obk,Guo:2010tk,Guo:2008zg}.
For the former, the conventional $5^3S_1$ charmonium state~\cite{Badalian:2008dv,Segovia:2008ta,Zhao:2023hxc} is a natural candidate.
For the latter, the structure observed in $e^+e^-\to\Lambda_c^+\bar{\Lambda}_c^-$ was first interpreted as a charmed baryonium state~\cite{Cotugno:2009ys}, i.e., a possible four-quark bound state in the charm sector. Further studies suggested that the $X(4630)$ and $\psi(4660)$ are the same state~\cite{Liu:2016sip}, and proposed various dynamical interpretations, such as a $\psi(2S)$-$f_0(980)$ bound state~\cite{Guo:2010tk,Guo:2008zg} or a $\Lambda_c^+\bar{\Lambda}_c^-$ bound state~\cite{Nakamura:2023obk}. Even with the explicit inclusion of a bare pole term, the $X(4630)$ and $\psi(4660)$ have been regarded as the same state~\cite{Dai:2017fwx}, stemming from the charmed baryon pair interaction.
However, none of these studies settles the key question of whether the $\psi(4660)$ pole originates from a renormalized charmonium state or from a purely dynamically generated state.

In this work, we address two central questions: whether the $\psi(4660)$ exists and, if so, what its nature is. These questions are of particular urgency because the Belle and BESIII datasets are mutually inconsistent in the $e^+e^-\to\Lambda_c^+\bar{\Lambda}_c^-$ channel; whether this discrepancy reflects different physics content or a common underlying structure that both measurements can constrain is precisely what our combined analysis aims to determine. Our strategy is as follows:

\begin{enumerate}[label=(\roman*), leftmargin=2.5em, labelsep=0.5em, align=left]
\item We construct, within $\sqrt{s}\in[4.0,\,5.0]~\mathrm{GeV}$ center-of-mass energy, an $S$-wave coupled-channel model comprising the open-charm baryon--antibaryon channels $\Lambda_c^+\bar{\Lambda}_c^-$, $\Xi_c^+\bar{\Xi}_c^-$, and $\Xi_c^0\bar{\Xi}_c^0$ (organized by SU(3) flavor symmetry), the $\sigma(500)/f_0(980)$-mediated $\psi(2S)\pi^+\pi^-$ channel, and the bare $4S$ and $5S$ vector charmonium states, and perform a combined fit to the Belle and BESIII data on an equal footing. As no $\Xi_c$ data are available yet, only the $\Lambda_c^+\bar{\Lambda}_c^-$ and $\psi(2S)\pi^+\pi^-$ cross sections enter the fit.
\item Following Refs.~\cite{Ye:2025ywy,Du:2016qcr}, we transform the hadronic basis into the SU(3) flavor singlet, octet, and isospin-triplet bases, and solve the Lippmann--Schwinger equation (LSE) with short-range contact potentials constrained by heavy-quark spin symmetry (HQSS); the physical cross sections are then obtained from the production amplitudes.
\item Using the parameters obtained from the fit, we search for the poles of the resonant states on the Riemann sheets. Decreasing the couplings to the bare charmonium states gradually and tracing the pole trajectories reveals whether the $\psi(4660)$ is a renormalized bare charmonia or dynamically generated states.
\item Moreover, the upcoming BEPCII upgrade~\cite{BESIII:2020nme} will scan the $4.95$--$5.6~\mathrm{GeV}$ energy region with high statistics, covering the ground-state charmed baryon thresholds; our predictions for the $e^+e^-\to\Xi_c\bar{\Xi}_c$ cross sections provide direct guidance for such experimental studies.
\end{enumerate}

A single framework that simultaneously describes both the Belle and the BESIII datasets would constitute, by itself, a quantitative arbitration of their apparent tension, and the $\Xi_c$ predictions presented here offer concrete benchmarks for the upcoming high-statistics BESIII measurements.

The paper is structured as follows. In Sec.~\ref{sec:Formalism}, we construct the coupled-channel formalism, using HQSS-constrained contact potentials and the LSE to obtain the cross sections for the relevant channels. In Sec.~\ref{sec:Results}, we perform a combined fit to the Belle and BESIII data, extract the pole positions and other physical quantities, and present our predictions for the $e^+e^-\to\Xi_c\bar{\Xi}_c$ cross section. A brief summary concludes the paper in Sec.~\ref{sec:Summary}.

\section{FORMALISM}\label{sec:Formalism}

We now construct the coupled-channel formalism in detail, following the strategy outlined in the Introduction.

\subsection{The contact potentials}

The light diquark inside the charmed baryon is treated as the $\bar{\mathbf{3}}$
conjugate representation of both SU(3) color and SU(3) flavor symmetries~\cite{Anselmino:1992vg,Shifman:2005wa,Kim:2024tbf,Niu:2026ssi}.
Because of Fermi statistics, the light diquark is realized as a compact spin-singlet configuration.
For instance, the $(ud)$ diquark in $\Lambda_c^+$ can be regarded as an effective antistrange quark $(\overline{s})$ in flavor space with spin zero.
This identification follows from the fact that the antisymmetric light diquark $(ud)$ transforms in the same $\bar{\mathbf{3}}$ representation as an antiquark under SU(3) flavor symmetry, and therefore shares identical transformation properties at the level of flavor symmetry.
Consequently, in the SU(3) symmetry limit, replacing the diquark by an effective antiquark provides a convenient and widely used approximation for organizing hadronic multiplets and constructing interaction vertices, although it should be understood as a representation-level equivalence rather than a dynamical one.
With this convention, the light quark components of the final-state charmed baryon pairs
$\Lambda_c^+\bar{\Lambda}_c^-\,,
\Xi_c^+\bar{\Xi}_c^-\,,
\Xi_c^0\bar{\Xi}_c^0\,$
correspond to
$s\bar{s}\,,d\bar{d}\,,u\bar{u}$, respectively.

Since the QED vertex is flavor-blind, electron-positron annihilation to charmed baryon pairs produces only SU(3) flavor states with total third component equal to zero. Accordingly, the zero-third-component SU(3) flavor basis can be constructed as
$|0\rangle=(u\bar{u}+d\bar{d}+s\bar{s})/\sqrt{3}$,
$|8\rangle=(u\bar{u}+d\bar{d}-2s\bar{s})/\sqrt{6}$, and
$|1\rangle=(u\bar{u}-d\bar{d})/\sqrt{2}$,
 which are orthogonal to each other, representing the SU(3) flavor singlet, octet and isospin triplet bases, respectively.
Within these bases, the charmed baryon pairs can be transformed from the hadronic basis into the SU(3) flavor basis as follows:
\begin{align}
    &|\Lambda_{c}^{+}\bar{\Lambda}^{-}_{c}\rangle^{0}=\frac{1}{\sqrt{3}}\left[|\Lambda_{c}^{+}\bar{\Lambda}^{-}_{c}\rangle+|\Xi_{c}^{+}\bar{\Xi}_{c}^{-}\rangle+|\Xi_{c}^{0}\,\bar{\Xi}_{c}^{0}\rangle\right] ~,\\
    &|\Lambda_{c}^{+}\bar{\Lambda}^{-}_{c}\rangle^{8}=\frac{1}{\sqrt{6}}\left[-2|\Lambda_{c}^{+}\bar{\Lambda}^{-}_{c}\rangle+|\Xi_{c}^{+}\bar{\Xi}_{c}^{-}\rangle+|\Xi_{c}^{0}\,\bar{\Xi}_{c}^{0}\rangle\right] ~,\\
    &|\Lambda_{c}^{+}\bar{\Lambda}^{-}_{c}\rangle^{1}=\frac{1}{\sqrt{2}}\left[-|\Xi_{c}^{+}\bar{\Xi}_{c}^{-}\rangle+|\Xi_{c}^{0}\,\bar{\Xi}_{c}^{0}\rangle\right] ~.
\end{align}
Here, $|\Lambda_{c}^{+}\bar{\Lambda}^{-}_{c}\rangle^{\gn}$ represents the SU(3) flavor basis, with $\gn=0,8,1$ denoting the SU(3) singlet, octet, and isospin triplet, respectively.

When the center-of-mass energy is as high as $4.6~\mathrm{GeV}$,
the pion invariant mass in the $\psi(2S)\pi^+\pi^-$ can reach up to $0.98~\mathrm{GeV}$, which
perfectly covers the $\sigma(500)$ and $f_0(980)$ resonance regions. In this case, the process can be described by two potential sequential processes
\begin{multline}
[e^+e^-\to \psi(2S)\sigma(500)\to \psi(2S)\pi^+\pi^-],
\end{multline}
and
\begin{multline}
e^+e^-\to \psi(2S)f_0(980)\to \psi(2S)\pi^+\pi^-.
\end{multline}
Accordingly, the two additional  channels,
$|\psi(2S)\sigma(500)\rangle$ and $|\psi(2S)f_0(980)\rangle$,
can be directly identified with the SU(3) flavor singlet basis, as follows:
\begin{align}
&|\psi'\sigma\rangle^{0}=|\psi(2S)\sigma(500)\rangle ~,\\
&|\psi'f\rangle^{0}=|\psi(2S)f_0(980)\rangle ~,
\end{align}
where the index 0 represents the SU(3) singlet. Finally, the full transformation between the hadronic basis and the SU(3) flavor symmetry basis reads
\begin{multline}
\left[
|\Lambda_{c}^{+}\bar{\Lambda}^{-}_{c}\rangle,
|\Xi_{c}^{+}\bar{\Xi}_{c}^{-}\rangle,
|\Xi_{c}^{0}\,\bar{\Xi}_{c}^{0}\rangle,
|\psi'\sigma\rangle,
|\psi'f\rangle
\right]^T= \\
R\left[
|\Lambda_{c}^{+}\bar{\Lambda}^{-}_{c}\rangle^{0},
|\Lambda_{c}^{+}\bar{\Lambda}^{-}_{c}\rangle^{8},
|\Lambda_{c}^{+}\bar{\Lambda}^{-}_{c}\rangle^{1},
|\psi'\sigma\rangle^{0},
|\psi'f\rangle^{0}
\right]^T,
\end{multline}
where the transformation matrix $R$ is
\begin{align}
R=
\begin{pmatrix}
\frac{1}{\sqrt{3}}&\frac{-2}{\sqrt{6}}&0&0&0\\
\frac{1}{\sqrt{3}}&\frac{1}{\sqrt{6}}&\frac{-1}{\sqrt{2}}&0&0\\
\frac{1}{\sqrt{3}}&\frac{1}{\sqrt{6}}&\frac{1}{\sqrt{2}}&0&0\\
0&0&0&1&0\\
0&0&0&0&1
\end{pmatrix} ~.
\label{Rmatrx}
\end{align}

The S-wave heavy-light decomposition of the SU(3) flavor representations introduced above reads~\cite{Voloshin:2012dk,Du:2016qcr,Ye:2025ywy}
\begin{multline}
|[s_{Q_1}s_{l_1}]_{j_1}[s_{Q_2}s_{l_2}]_{j_2}\rangle_J=
\sum_{s_l,s_Q}\hat{s_Q}\,\hat{s_l}\,\hat{j_1}\,\hat{j_2}\\
\times\left\{
\begin{array}{@{}ccc@{}}
s_{Q_1} & s_{Q_2} & s_Q \\
s_{l_1} & s_{l_2} & s_l \\
j_1 & j_2 & J
\end{array}
\right\}
|[s_{Q_1}s_{Q_2}]_{s_Q}[s_{l_1}s_{l_2}]_{s_l}\rangle_J,
\label{1}
\end{multline}
with $\hat{j}=\sqrt{2j+1}$.
The $|[s_{Q_1}s_{l_1}]_{j_1}[s_{Q_2}s_{l_2}]_{j_2}\rangle_J$
represents SU(3) flavor basis for a pair of charmed baryons,
with $s_{Q_i}$ ($s_{l_i}$) the spin of the heavy quark (the light degree of freedom which is the sum of light quark spin and the relative orbital angular momentum) inside the $i$th hadron. The total spin of the $i$th hadron is given by $j_i$, and $J$ stands for the total angular momentum of the two-hadron system.
From the $|[s_{Q_1}s_{Q_2}]_{s_Q}[s_{l_1}s_{l_2}]_{s_l}\rangle_J$ basis, one can read the heavy and light degrees of freedom, separately. $s_Q$ denotes the total spin of the heavy quark pair, while $s_l$ represents the total angular momentum of the light degrees of freedom, including both the total spin of the light quarks and the relative orbital angular momentum between the two hadrons.

In the heavy quark limit, the $s_Q$ and $s_l$ are conserved separately.
Therefore, it is convenient to transform the $|[s_{Q_1}s_{Q_2}]_{s_Q}[s_{l_1}s_{l_2}]_{s_l}\rangle_J$  basis into the heavy-light basis  $|s_Q\otimes s_l\rangle_J$.
The $S$-wave states in the SU(3) flavor basis are given by
\begin{align}
|\Lambda_{c}^{+}\bar{\Lambda}^{-}_{c}\rangle_{1^{--}}^{\gn}=|1\otimes 0\rangle^{\gn}\quad({\gn}=0,8,1).
\end{align}
The according low-energy constants (LECs) can be defined as
\begin{align}
C_{\gn}\equiv{}^{\gn}\langle1\otimes0|\mathcal{H}_{CT}|1\otimes0\rangle^{\gm}\delta^{\gn\gm}\quad({\gn,\gm}=0,8,1),
\label{2}
\end{align}
where repeated indices are not summed over. Here, $\mathcal{H}_{CT}$ denotes the leading-order contact Hamiltonian that respects HQSS. As we focus on the energy region $[4.0, 5.0]~\mathrm{GeV}$, the contact potentials $C_{\gn}$ are treated as energy-independent constants. The remaining two reaction channels are defined as follows:
\begin{align}
C_{\gp}\equiv{}^{0}\langle1\otimes 0|\mathcal{H}_{CT}|1\otimes 0\rangle^{0}_{\gp} ~,
\label{3}
\end{align}
where the subscript ${\gp}=\psi'\sigma,~\psi' f$ in $C_{\gp}$ and $|1\otimes 0\rangle^{0}_{\gp}$ refers to the two channels $\psi(2S)\sigma(500)$ and $\psi(2S)f_0(980)$, respectively. Both channels belong to the SU(3) flavor singlet.
According to Eqs.~(\ref{2})--(\ref{3}), the contact potential matrix for the five channels in the SU(3) flavor basis can be written as follows:
\begin{align}
V^{\rm flav}_{ij} =
\begin{pmatrix}
C_0 & 0 & 0 &C_{\psi'\sigma} &C_{\psi'f}\\
0 & C_8 & 0 &0&0 \\
0 & 0 & C_1 &0&0 \\
C_{\psi'\sigma} & 0 &0 &0&0 \\
C_{\psi'f} & 0 &0 &0&0
\end{pmatrix}.
\label{Voo_flav}
\end{align}
In the following discussion, the Latin subscripts $i,j,m,n=1,\dots,5$ denote the five channels
$|\Lambda_c^+\bar{\Lambda}_c^-\rangle$,
$|\Xi_c^+\bar{\Xi}_c^-\rangle$,
$|\Xi_c^0\bar{\Xi}_c^0\rangle$,
$|\psi'\sigma\rangle$, and
$|\psi'f\rangle$, respectively.
Within the energy region of interest $[4.0, 5.0]~\mathrm{GeV}$, there are also two additional conventional charmonia, $\psi(4S)$ and $\psi(5S)$, which are SU(3) flavor singlet and labeled by the Greek indices $\alpha,\beta,\gamma=1,2$.
 Consequently, their heavy-light basis,
$|1\otimes 0 \rangle^{0}_{4S}$ and $|1\otimes 0 \rangle^{0}_{5S}$,
exclusively couple to the $|1\otimes 0 \rangle^{0}$ component.
The coupling constants $g_{4S}$ and $g_{5S}$ are defined through the Hamiltonian density $\mathcal{H}_{\rm bare}$, which describes the interaction between the bare charmonium states and the channels labeled by $i=1,\dots,5$, as follows:
\begin{align}
g^0_{4S} &\equiv {}^{0}\langle 1\otimes 0 | \mathcal{H}_{bare} | 1\otimes 0 \rangle^{0}_{4S} ~,\label{4}\\
g^0_{5S} &\equiv {}^{0}\langle 1\otimes 0 | \mathcal{H}_{bare} | 1\otimes 0 \rangle^{0}_{5S} ~.
\label{5}
\end{align}
With these definitions, the constant potential of two hidden-charm channels in the SU(3) flavor basis can then be written as:
\begin{align}
V^{\rm flav}_{i\alpha} =
\begin{pmatrix}
g^0_{4S} & 0 & 0 & 0 & 0 \\
g^0_{5S} & 0 & 0 & 0 & 0
\end{pmatrix}^T ~.
\label{Vob_flav}
\end{align}

\subsection{The Lippmann-Schwinger equation and the physical production amplitude}
At a given center-of-mass (c.m.) energy $E$, the $T$-matrix is obtained by solving the Lippmann-Schwinger equation (LSE)
\begin{align}
T(E) = V +VG(E)T(E).
\label{LSE}
\end{align}
Here, $V$ denotes the corresponding potential that encompasses all seven channels, whose explicit form reads
\begin{align}
V =
\begin{pmatrix}
[V_{ij}]_{5\times5} & [V_{i\alpha}]_{5\times2}  \\
[V_{\beta j}]_{2\times5} & [\bm{0}]_{2\times2}
\end{pmatrix},
\label{Vmatrix}
\end{align}
and $G(E)$ represents the two-point loop function matrix, which can be written in the form~\cite{Guo:2015umn}
\begin{align}
G(E) =
\begin{pmatrix}
[G_{ij}(E)]_{5\times5} &  [\bm{0}]_{5\times2} \\
[\bm{0}]_{2\times5} & [G_{\alpha\beta}(E)]_{2\times2}
\end{pmatrix},
\label{Gmatrix}
\end{align}
where the two-body propagator of the $i$th channel is
\begin{align}
&G_{ii}(E)=\frac{1}{4m_{i1}m_{i2}}\int\frac{d^3q}{(2\pi)^3}\frac{f_{\Lambda}^2({\vec{q}}^{\,2})}{E-m_{i1}-m_{i2}-{\vec{q}}^{\,2}/(2\mu)} \notag\\
&=\frac{1}{4m_{i1}m_{i2}}\left\{\frac{-\mu\Lambda}{(2\pi)^{3/2}}+\frac{\mu k}{2\pi}e^{\frac{-2k^2}{{\Lambda}^2}}\left[\text{erfi}\left(\frac{\sqrt{2}k}{\Lambda}\right)-\iu\right]\right\},
\label{2PF}
\end{align}
with $m_{i1}$ and $m_{i2}$ the masses of the two hadrons in the $i$th channel and $\mu=\frac{m_{i1}m_{i2}}{m_{i1}+m_{i2}}$ the corresponding reduced mass. $k = \sqrt{2 \mu (E - m_{i1} - m_{i2})}$ is the on-shell center-of-mass three-momentum. Note that $G_{ij}\equiv0$ for $i\neq j$, and the bold symbol $\iu $ denotes the imaginary unit, to be distinguished from the channel index $i$. Here we use the Gaussian form factor
$f_{\Lambda}(q^2)=\text{exp}(-q^2/\Lambda^2)$ to regularize the divergence in the two-body propagator. The bare-pole propagator is diagonal as well, with the $\alpha$th
 pole expressed as $G_{\alpha\alpha}(E)=(E^2-m_{\alpha}^2+\iu\epsilon)^{-1}$.

To describe the contributions $\sigma(500)/f_0(980)$ in the $\psi(2S)-\sigma(500)/f_0(980)$ two-point function, the propagator of $\sigma(500)/f_0(980)$ in the loop integral is replaced by its Flatté parametrization~\cite{Flatte:1976xu}~\footnote{Although the Flatté parametrization does not describe the complicated $\sigma(500)$ line shape exactly~\cite{Giacosa:2007bn,Tornqvist:1995ay,Achasov:2010kk}, it is adequate for our purpose, since the line shapes of the intermediate states are not the focus of this work. For simplicity, we therefore adopt the same parametrization for both the $\sigma(500)$ and the $f_0(980)$, with the difference that the $\sigma(500)$ parametrization does not include the $K\bar{K}$ channel.} For concreteness, we present the explicit expression for the case $\psi(2S)-f_0(980)$ as an example:
\begin{multline}
G_{\psi'f}=\iu\int\frac{d^4q}{(2\pi)^4}f(|\Vec{q}|^2) [(p-q)^2-m_{\psi}^2+\iu\epsilon]^{-1}\\
\times~[q^2-m_f^2+\iu m_f(\Gamma_{\pi\pi}+\Gamma_{K\Bar{K}})]^{-1},
\label{Flatte2PF}
\end{multline}
with relativistic partial widths
\begin{align}
\Gamma^f_{\pi\pi} &= \bar{g}_{f\pi} \sqrt{q^2/4-m_{\pi}^2} ,\label{PartialWidthsPi}\\
\Gamma^f_{K\Bar{K}} &=
\begin{cases}
\bar{g}_{f_K} \sqrt{q^2/4-m_{K}^2} & \quad q^2>4m_{K}^2 \\[1ex]
i\,\bar{g}_{f_K} \sqrt{m_{K}^2-q^2/4} & \quad q^2<4m_{K}^2
\end{cases}.
\label{PartialWidthsK}
\end{align}
Although the energy-dependent width in Eq.~\eqref{PartialWidthsPi} is not a typical Breit-Wigner parametrization of the $\sigma(500)$, it is one of the parametrizations used in the BESIII analysis on the $\psi(2S)\to J/\psi\pi^+\pi^-$ process~\cite{BES:2006eer}, as well as in the analysis of Ref.~\cite{E791:2000vek}. In Ref.~\cite{BES:2006eer}, four parametrization schemes are used to describe the $\pi\pi$ channel, i.e. the standard Breit-Wigner constant width, a constant width plus a kinematic factor, the P.K.U. parametrization which removes the spurious singularity in the second parametrization, and the parametrization with explicit Adler zero. In our framework, we use the second form (similar to a Flatt\'e-type parametrization). Furthermore, to check its difference from the standard Breit-Wigner parametrization, we perform an explicit cross-check by replacing the energy-dependent width with a constant Breit--Wigner width ($\Gamma^{\sigma}=558$~MeV, as reported in Ref.~\cite{Hoferichter:2023mgy}) for three different values of the cutoffs ($\Lambda=3.4$, $3.8$, and $4.2$~GeV). By repeating the fit, we find that the resulting line shapes and pole positions are qualitatively consistent. The detailed comparisons are presented in
Appendix~\ref{AAconstBW}.
Here, dimensionless couplings $\bar{g}_{f\pi}$ and $\bar{g}_{f_K}$ are related to commonly used dimensional couplings $g_{f\pi\pi}$ and $g_{f_{K\bar{K}}}$ via
$\Bar{g}_{f\pi}=g_{f\pi\pi}^2/(8\pi m_{f}^2)$
and $\Bar{g}_{f_K}=g_{f_{K\bar{K}}}^2/(8\pi m_{f}^2)$, where $m_f$ is the mass of the resonance~\cite{Baru:2004xg}. Despite replacement with Flatté parametrization, the loop function $G_{\psi'f}$ can still be expressed in a nonrelativistic analytic form analogous to Eq.~(\ref{2PF}). Detailed derivation steps are provided in the Appendix~\ref{AAfla2PF}. Consequently, for the two-point function $\psi(2S)-f_0(980)$, the only modification required is to replace $k$ in Eq.~(\ref{2PF}) with
\begin{align}
k=\sqrt{2\mu(E-m_{\psi}-m_f+\iu\Gamma^f_{\rm tot}/2)},
\label{flatteK}
\end{align}
with $\mu$ the reduced mass of the $\psi(2S)$--$f_0(980)$ system.
Here, $\Gamma^f_{\rm tot}$ denotes the total nonrelativistic width, defined as the sum of the partial widths:
\begin{multline}
\Gamma^f_{\rm tot}=\bar{g}_{f\pi}\sqrt{m_\pi(E-m_\psi-2m_\pi)}\\+\bar{g}_{f_K}\sqrt{m_K(E-m_\psi-2m_K)}.
\end{multline}

\begin{figure}[H]
\centering
\includegraphics[width=0.95\columnwidth]{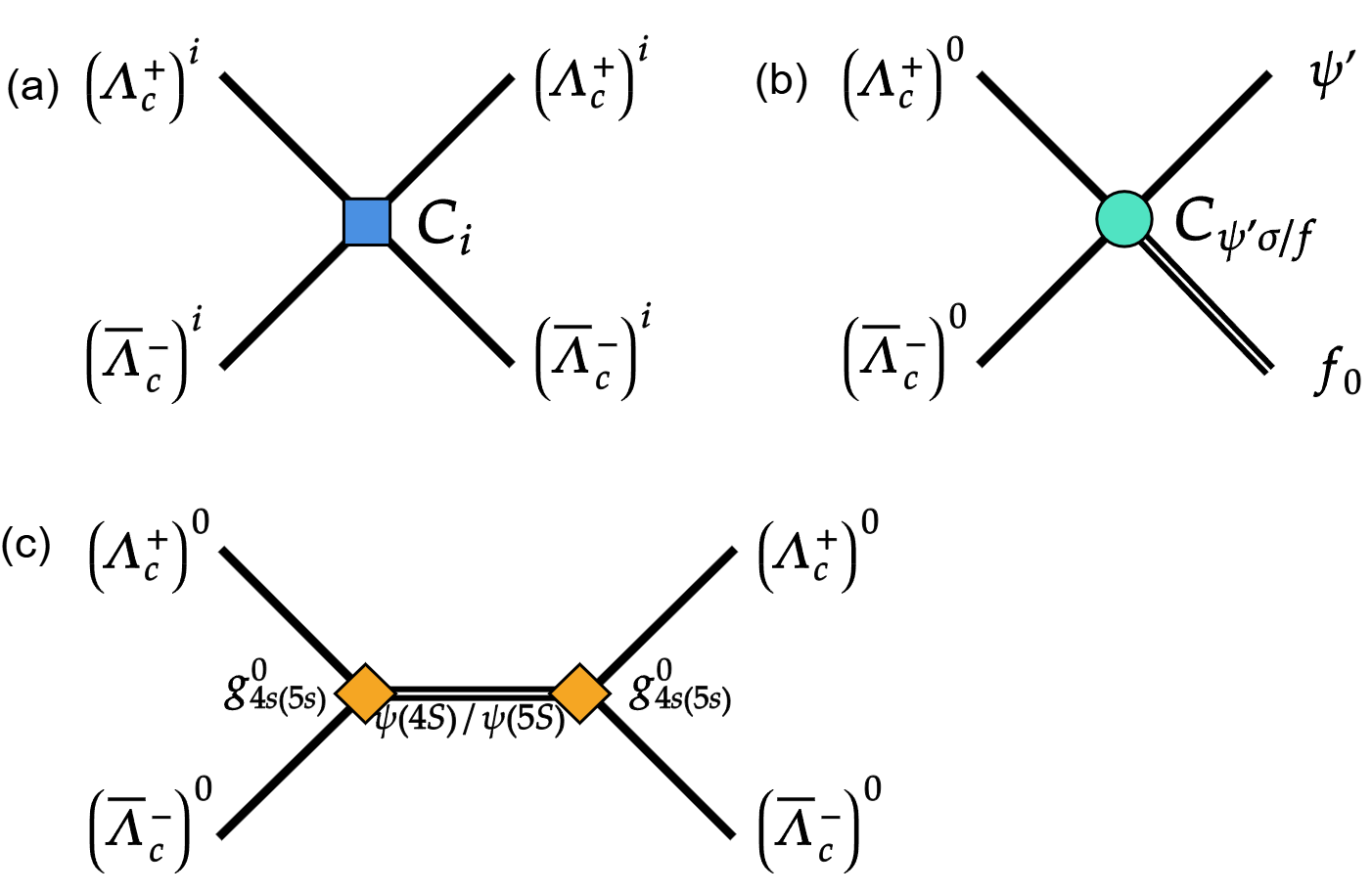}
\caption{(a)(b): Short-range contact potentials. (c): Effective potential mediated by the bare resonance states $\psi(4S)$ or $\psi(5S)$.}
\label{fig:Cont}
\end{figure}

Substituting Eqs.~(\ref{Vmatrix})--(\ref{Gmatrix}) into Eq.~(\ref{LSE}), one can obtain
\begin{widetext}
\begin{align}
\begin{pmatrix}
[T_{ij}]_{5\times5} & [T_{i\alpha}]_{5\times2}  \\
[T_{\beta j}]_{2\times5} & [T_{\beta\alpha}]_{2\times2}
\end{pmatrix}=
\begin{pmatrix}
[V_{ij} + V_{im}\, G_{mn}\, T_{nj} + V_{i\alpha}\, G_{\alpha\beta}\, T_{\beta j}]_{5\times5} &\,
[V_{i\alpha} + V_{im}\, G_{mn}\, T_{n\alpha} + V_{i\beta}\, G_{\beta\gamma}\, T_{\gamma\alpha}]_{5\times2} \\
[V_{\beta j} + V_{\beta m}\, G_{mn}\, T_{nj}]_{2\times5} &\,
[V_{\beta i}\, G_{ij}\, T_{j\alpha}]_{2\times2}
\end{pmatrix}.
\end{align}
\end{widetext}
Here, $T_{ij}$ denotes the scattering amplitude between the five channels labeled with $i,j=1,2,3,4,5$, $T_{i\alpha}$ represents the scattering amplitude from the charmonium state to the channel $i$, and $T_{\alpha i}$ corresponds to the inverse process of $T_{i\alpha}$.
Subsequently, plugging $T_{\beta j}$ into $T_{ij}$ obtains the scattering amplitudes among the five channels labeled by $i=1,\dots,5$ in terms of the effective potential:
\begin{align}
T_{ij}(E) = \hat{V}_{ij}^{\rm eff}(E) + \hat{V}_{im}^{\rm eff}(E)\, G_{mn}(E)\, T_{nj}(E),
\label{effLSE}
\end{align}
where the effective potential is defined as~\cite{Hanhart:2015cua,Guo:2016bjq,Baru:2010ww,Hanhart:2011jz}
\begin{align}
\hat{V}_{ij}^{\rm eff}(E) \equiv V_{ij} + V_{i\alpha}\, G_{\alpha\beta}(E)\, V_{\beta j}
\end{align}
with its schematic diagram presented in Fig.~\ref{fig:Cont}.
In our case, the effective potential in the hadronic basis is obtained via the following basis transformation:
\begin{align}
\hat{V}_{ij}^{\rm eff}(E)
&= R\Big(V^{\rm flav}_{ij} + V^{\rm flav}_{i\alpha}\, G_{\alpha\beta}(E)\, {V^{\rm flav}_{\beta j}}^T \Big)R^{-1} \notag\\
&=R\hat{V}_{ij}^{\rm eff'}(E)R^{-1},
\end{align}
with $\hat{V}_{ij}^{\rm eff'}(E)$ the effective potential based on the SU(3) flavor basis and $R$ the transformation matrix.
And $\hat{V}_{ij}^{\rm eff'}(E)$ can
be written as follows:
\begin{align}
\hat{V}_{ij}^{\rm eff'}(E) =
\begin{pmatrix}
C_{\rm co}(E) & 0 & 0 &C_{\psi'\sigma} &C_{\psi'f}\\
0 & C_8 & 0 &0&0 \\
0 & 0 & C_1 &0&0 \\
C_{\psi'\sigma} & 0 &0 &0&0 \\
C_{\psi'f} & 0 &0 &0&0
\end{pmatrix},
\label{Voo_flav_eff}
\end{align}
with $C_{\rm co}(E)=C_0+(g^0_{4S})^2 G_{4S}(E)+(g^0_{5S})^2 G_{5S}(E)$.
Ultimately, the full $T$-matrix for the coupled-channel system can be obtained by solving Eq.~(\ref{effLSE})
\begin{align}
T_{ij}(E) = f_{\Lambda}(p) \left[ [\hat{V}_{ij}^{\rm eff}(E)]^{-1} - G_{ij}(E) \right]^{-1} f_{\Lambda}(p').
\label{Too}
\end{align}
To preserve unitarity of the $T$-matrix, the Gaussian form factor $f_\Lambda(p)$ originating from the loop function $G_{ij}(E)$, is explicitly included in the expression of the $T$-matrix above.

The bare production amplitude is defined as
\begin{align}
\mathcal{F} = ([F_{i}]_{5\times1}^T,\, [f_{\alpha}]_{2\times1}^T)^T,
\label{Fmatrix}
\end{align}
where $F_i$ denotes the bare production amplitude between the virtual photon and the five channels$(i=1,\dots,5)$ in the SU(3) flavor basis, and $f_\alpha$ represents the bare production amplitude between the virtual photon and the charmonium states.
The physical production amplitude can be obtained from
\begin{align}
\mathcal{U}(E) = \mathcal{F} + V\, G(E)\, \mathcal{U}(E).
\label{LSE_U}
\end{align}
Substituting Eqs.~(\ref{Vmatrix})--(\ref{Gmatrix}) and Eq.~(\ref{Fmatrix}) into $\mathcal{U}(E)$, one can obtain the physical production amplitudes
\begin{align}
&[\mathcal{U}_{i}]_{5\times1} = F_i + V_{ij} G_{jm}(E) \mathcal{U}_{m} + V_{i\alpha} G_{\alpha\beta}(E) \mathcal{U}_{\beta} \label{Uo},\\
&[\mathcal{U}_{\alpha}]_{2\times1} = f_\alpha + V_{\alpha i} G_{ij}(E) \mathcal{U}_{j}.
\label{Ub}
\end{align}
Subsequently, by substituting $\mathcal{U}_{\alpha}$ into $\mathcal{U}_{i}$, we obtain the physical production amplitudes for the five channels $(i=1,\dots,5)$ in terms of the effective potential:
\begin{align}
\mathcal{U}_{i}(E) = f_{\Lambda}(p) \Big(\bm{1}_{5\times5} - \hat{V}_{ij}^{\rm eff}(E)\, G_{jm}(E)\Big) ^{-1} \hat{F}_{m}^{\rm eff}(E).
\label{EffU}
\end{align}
Here, $\hat{F}_{i}^{\rm eff}(E) \equiv F_{i} + V_{i\alpha}\, G_{\alpha\beta}\, f_{\beta} $
represents the effective production amplitude, and the Gaussian form factor $f_{\Lambda}(p)$ is introduced analogously to the $T$-matrix part in order to regularize the ultraviolet divergence in the loop integral.
Analogous to the effective potential, the effective bare production amplitude in SU(3) flavor basis is given by
\begin{align}
\hat{F}_{i}^{\rm eff'}(E) = F_{i}^{\rm flav} + V^{\rm flav}_{i\alpha}\, G_{\alpha}(E)\, f_{\alpha}^{\rm flav}.
\end{align}
where $F_{i}^{\rm flav} = (F^0,F^8,F^1,0,0)^T$
and $f_{\alpha}^{\rm flav} = (f_{4S}^0,f_{5S}^0)^T$.
The coupling between the virtual photon and the baryon-antibaryon channel $F^{\gn}\,({\gn}=0,8,1)$ is given by $F^{\gn} \equiv f_S^{\gn} + f_D^{\gn}$,
where $f_S^{\gn}$ and $f_D^{\gn}$ denote the couplings between the photon and the light degrees of freedom in corresponding channel for the $s_l=0~{\rm and}~2$ components, respectively.
With the transformation matrix $R$,
\begin{align}
\hat{F}_{i}^{\rm eff}(E) = R~\hat{F}_{i}^{\rm eff'}(E),
\end{align}
one derives the effective production amplitudes expressed in the hadronic basis.

\subsection{The cross section}\label{sec:FormalismC}

\begin{figure}[tb]
\centering
\includegraphics[width=0.8\columnwidth]{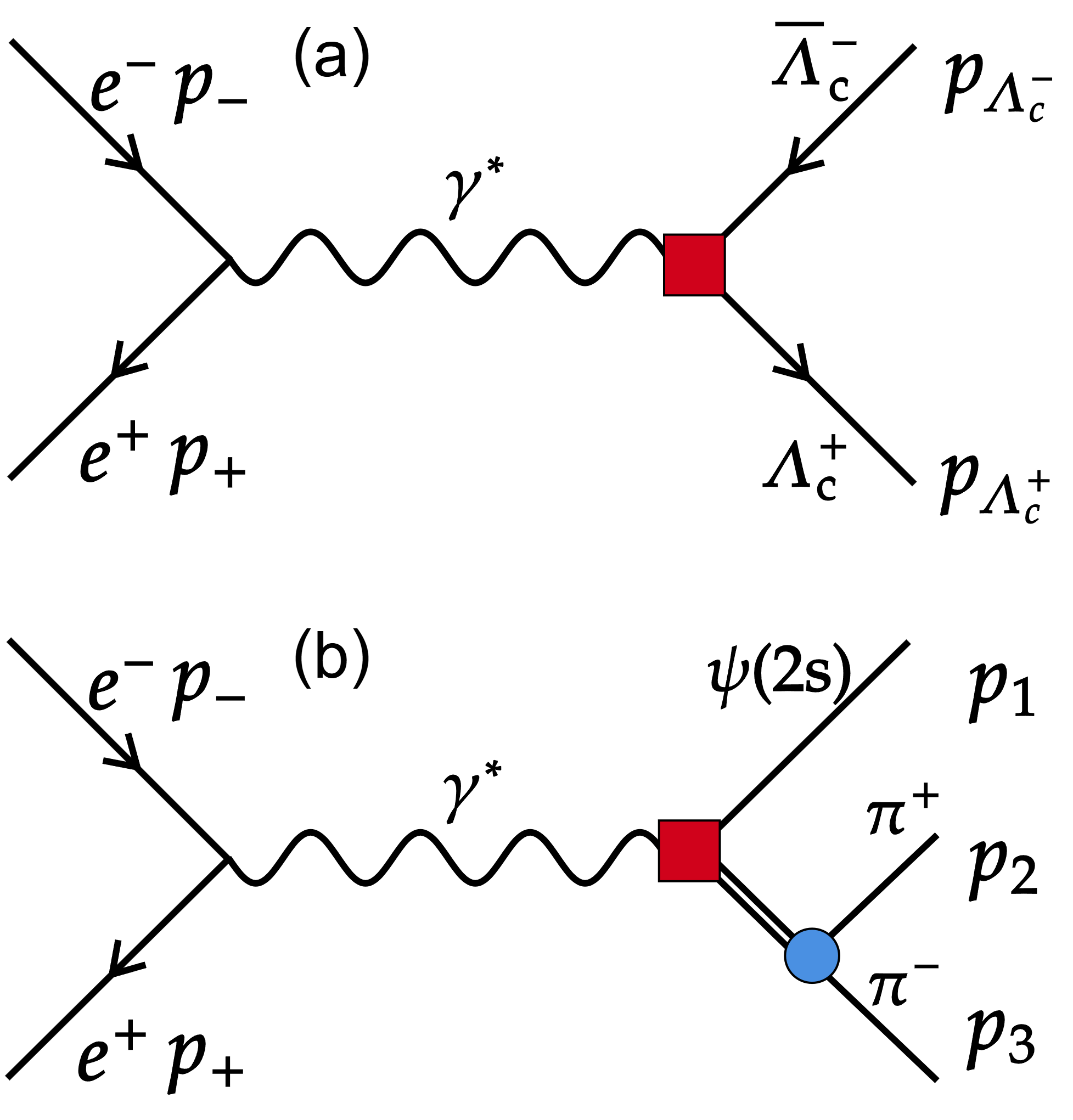}
\caption{Feynman diagrams for $e^+e^-\rightarrow\Lambda^+_c\bar{\Lambda}^-_c$ and $e^+e^-\rightarrow\psi(2S)\,\sigma(500)/f_0(980)\rightarrow\psi(2S)\pi^+\pi^-$.
Red square denotes the physical production amplitude and blue circle denotes the coupling between the $\pi\pi$ system and the intermediate resonance.}
\label{fig:FeynD}
\end{figure}

The scattering amplitudes for the three processes depicted in Fig.~\ref{fig:FeynD} are given by
\begin{multline}
\mathcal{M}\Big(e^+e^-\to \Lambda_{c}^{+}\bar{\Lambda}^{-}_{c}\Big)=
\Big[\bar{v}(p_+)(-{\iu}e\gamma^{\mu})u(p_-)\Big] \\
\times\left(\frac{-{\iu}g_{\mu\nu}}{s}\right)\bar{u}(p_{{\Lambda}^{-}_{c}})\left(-{\iu}e\, \mathcal{U}_1\gamma^{\nu}\right)v(p_{{\Lambda}^{+}_{c}}),\label{M11}
\end{multline}
\begin{multline}
\mathcal{M}\Big(e^+e^-\to \psi(2S)\sigma(500)\rightarrow\psi(2S)\pi^+\pi^-\Big)\\
=\Big[\bar{v}(p_+)(-{\iu}e\gamma^{\mu})u(p_-)\Big]\left(\frac{-{\iu}g_{\mu\nu}}{s}\right)\mathcal{U}_{4}\,f_{\rm el}^{\sigma}\,\epsilon^{*\nu},\label{M44}
\end{multline}
\begin{multline}
\mathcal{M}\Big(e^+e^-\to \psi(2S)f_0(980)\rightarrow\psi(2S)\pi^+\pi^-\Big)\\
=\Big[\bar{v}(p_+)(-{\iu}e\gamma^{\mu})u(p_-)\Big]\left(\frac{-{\iu}g_{\mu\nu}}{s}\right)\mathcal{U}_{5}\,f_{\rm el}^{f}\,\epsilon^{*\nu},\label{M55}
\end{multline}
with $\mu,\nu$ Lorentz indices, $s$ the squared center-of-mass energy, and $p_{\Lambda_c^\pm}$ the four-momentum of $\Lambda_c^+ (\bar{\Lambda}_c^-)$.
Here, $\epsilon^*$ denotes the polarization vector of $\psi(2S)$ and obeys the relation $\epsilon^*_\mu\epsilon_\nu = -g_{\mu\nu}+\frac{p_{1\mu}p_{1\nu}}{(p_1)^2}$.
$\mathcal{U}_i\,(i=1,\dots,5)$ represent the physical production amplitudes and correspond respectively to the five matrix elements derived from Eq.~(\ref{EffU}).
We parametrize the final-state processes
$\sigma(500) \to \pi\pi$ and $f_0(980) \to \pi\pi$, respectively,
using the Flatt\'e form as $f^\sigma_{\rm el}$ and $f^f_{\rm el}$~\cite{Flatte:1976xu,Baru:2004xg}, which are given by
\begin{align}
&f_{\rm el}^{\sigma}=\dfrac{m_{\sigma}g_{\sigma\pi\pi}}{m_{\pi\pi}^2-m_{\sigma}^2+{\iu}m_{\sigma}\Gamma^{\sigma}_{\pi\pi}}~,\label{fla500}\\
&f_{\rm el}^{f}=\dfrac{m_fg_{f\pi\pi}}{m_{\pi\pi}^2-m_f^2+{\iu}m_f(\Gamma^{f}_{\pi\pi}+\Gamma^f_{K\bar{K}})}~.\label{fla980}
\end{align}
Here, $\Gamma^{\sigma}_{\pi\pi}$, $\Gamma^f_{\pi\pi}$ and $\Gamma^f_{K\bar{K}}$ represent the partial widths in the Flatt\'e parametrization, with $m_{\pi\pi}^2=s_2$ being the squared invariant mass of the $\pi\pi$ system.

From Eq.~(\ref{M11}), the differential cross section for $e^+e^-\to \Lambda_{c}^{+}\bar{\Lambda}^{-}_{c}$ can be expressed as
\begin{align}
\left(\frac{d\sigma_1}{d\Omega}\right)_{\rm CM} =
\frac{\alpha^2\,\mathcal{U}_1^2}{ s^{3/2}}|\Vec{p}_{{\Lambda}^{-}_{c}}|~.
\end{align}
where $\alpha=\frac{1}{137}$ is the fine structure constant. By performing the full phase-space integration, we subsequently obtain the scattering cross section as
\begin{align}
\sigma_1 = \frac{4\pi\alpha^2\,\mathcal{U}_1^2}{ s^{3/2}}|\Vec{p}_{{\Lambda}^{-}_{c}}|~.\label{sigma1}
\end{align}
Given the high precision of the BESIII energy scan data, we explicitly incorporate the Sommerfeld factor (SF)~\cite{Sommerfeld:1931qaf,Sakharov:1948plh} into our model by multiplying Eq.~(\ref{sigma1}) by a factor
\begin{align}
\frac{\pi\alpha}{\beta}\frac{\sqrt{1-\beta^2}}{1-{\rm exp}(-\pi\alpha\sqrt{1-\beta^2}/\beta)},
\end{align}
where $\beta=\sqrt{1-4M^2_{\Lambda_c^+}/s}$ is the velocity of the $\Lambda_c^+$ baryon.
This is essential to account for the strong Coulomb interaction between the charged pair $\Lambda_c^+\bar{\Lambda}_c^-$, which is responsible for the nonzero cross section observed just 1.6 MeV above the threshold~\cite{BESIII:2017kqg}.

The differential cross sections, $d\sigma_4$ and $d\sigma_5$, for the two processes illustrated in Fig.~\ref{fig:FeynD}(b) are given by
\begin{align}
d\sigma=\frac{1}{256 \pi^3 s^2}\, \frac{\overline{|\mathcal{M}|}^2}{(s_2-t_1)}dt_1\,dt_2\,ds_2 ~,
\end{align}
where $\overline{|\mathcal{M}|}^2$ is the squared amplitude, given in Eqs.~(\ref{M44})--(\ref{M55}) , averaged over initial spins and summed over final polarizations.
Three invariants are given in the c.m. frame by
\begin{align}
t_1=(p_--p_1)^2,
t_2=(p_+-p_3)^2,
s_2=(p_2+p_3)^2.
\end{align}
Detailed steps are provided in the Appendix~\ref{AA3body}.
The total cross section for $e^+e^-\to\psi(2S)\pi^+\pi^-$ is obtained by summing $\sigma_4$ and $\sigma_5$.

 \section{RESULTS AND DISCUSSION}\label{sec:Results}

In this section, we perform a combined fit to the experimental cross sections for the $e^+e^-\to\Lambda^+_c\bar{\Lambda}^-_c$~\cite{Belle:2008xmh,BESIII:2017kqg,BESIII:2023rwv} and $e^+e^-\to\psi(2S)\pi^+\pi^-$~\cite{Belle:2014wyt,BESIII:2021njb} processes from both Belle and BESIII collaborations.
Both datasets are treated on an equal footing. To make the fit converge quickly, we switch on the $0$ channel and $8$ channel step by step, which gives reasonable parameters as well. The detailed fitting steps are as follows:
\begin{itemize}
\item We first switch off the $8$ channel (i.e., fixed $C_8 = f_S^8 = 0$) and perform a fit using the $0$ channel alone, adjusting the parameters until the line shape and the $\chi^2$ reached their optimum.
\item Once the optimum was reached, we fixed the $0$ channel and switched on the $8$ channel, allowing it to enter in the previous fit.
\item At the end, we release all the parameters and fit again.
\end{itemize}

\begin{figure*}[!htbp]
\centering
\includegraphics[width=1\textwidth]{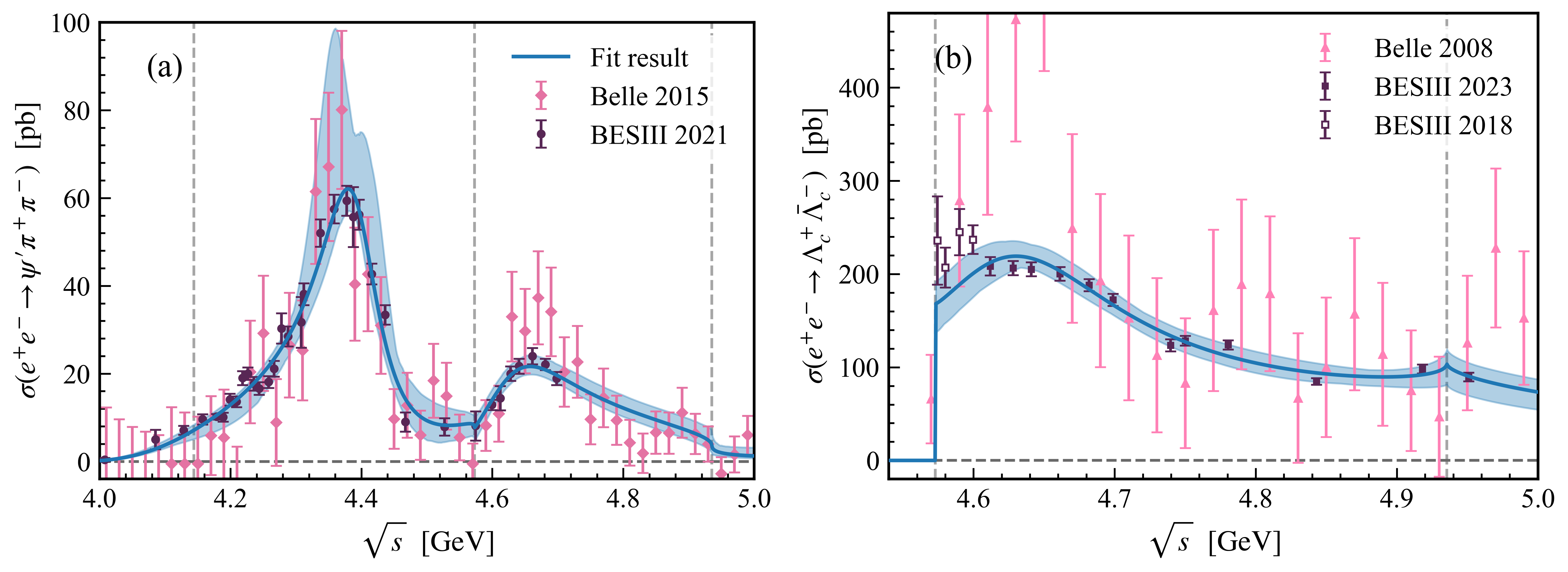}
\caption{
Comparison of the fitted line shapes (blue solid line) with experimental measurements for (a) $e^+e^-\to\psi'\pi^+\pi^-$ and (b) $e^+e^-\to\Lambda^+_c\bar{\Lambda}^-_c$. Data are from both Belle~\cite{Belle:2008xmh,Belle:2014wyt} (pink triangles and diamonds) and BESIII~\cite{BESIII:2017kqg,BESIII:2021njb,BESIII:2023rwv} (purple circles, purple boxes and hollow squares) collaborations.
The vertical dashed lines indicate the relevant thresholds, from left to right:
$\psi(2S)\sigma(500)$,
$\Lambda_c^+\bar{\Lambda}_c^-$, and
$\Xi_c\bar{\Xi}_c$.
}
\label{fig:FitResult}
\end{figure*}

The fitted parameters are presented in Tab.~\ref{parameters}, and the comparison between the fitted curves and the experimental data is shown in Fig.~\ref{fig:FitResult}.
The fitted masses for the $4^3S_1$ and $5^3S_1$ charmonium states are in good agreement with those reported in Refs.~\cite{Badalian:2008dv,Segovia:2008ta,Zhao:2023hxc}. The extracted interaction parameters are of the same order of magnitude as expected and fall within an acceptable range, while the remaining parameters are found to be of natural size.
Since the mass difference between the charged and neutral $\Xi_c$ baryons is only about $2.73~\mathrm{MeV}$, the thresholds of $\Xi_c^+\bar{\Xi}_c^-$ and $\Xi_c^0\bar{\Xi}_c^0$ are extremely close to each other.
Consequently, the associated Riemann sheets are within a narrow region. Given that the peak structures of interest in this work lie well below both thresholds, their impact on the near-threshold dynamics is expected to be negligible.
Accordingly, we work in the isospin-symmetric limit, i.e. setting the $\Xi_c^+$ and $\Xi_c^0$ masses equal to each other, treating the two thresholds as degenerate. The SU(3) flavor symmetry breaking effect comes from the mass difference between $\Lambda_c$ and $\Xi_c$.

\begin{table}[!htbp]
\centering
\setlength{\tabcolsep}{15pt} 
\renewcommand{\arraystretch}{1.3}
\caption{The parameters obtained from the combined fit to both Belle and BESIII experimental data.}
\label{parameters}
\begin{tabular}{l r@{\,}c@{\,}l}
\midrule\midrule
\textbf{Parameters} & \multicolumn{3}{c}{\textbf{Values}} \\
\cmidrule(lr){1-4}
$g^0_{4S}$ [GeV$^{0}$]
& -4.73   & $\pm$ & 0.30  \\
$g^0_{5S}$ [GeV$^{0}$]
&  9.89   & $\pm$ & 0.49  \\
$m_{4S}$ [MeV]
&  4386.99  & $\pm$ & 17.10  \\
$m_{5S}$ [MeV]
&  4611.52  & $\pm$ &  7.77  \\
$f^0_{4S}$ [$10^{-1}$GeV$^{2}$]
&   0.47    & $\pm$ & 0.03 \\
$f^0_{5S}$ [$10^{-1}$GeV$^{2}$]
&   0.46    & $\pm$ & 0.02 \\
$f_S^0$ [$10^{-1}$GeV$^{0}$]
&   1.46    & $\pm$ & 0.53 \\
$f_S^8$ [$10^{-1}$GeV$^{0}$]
&   -5.18   & $\pm$ & 0.28 \\
$C_0$ [GeV$^{-2}$]
&   91.18   & $\pm$ & 17.85 \\
$C_8$ [GeV$^{-2}$]
&  -58.74   & $\pm$ & 2.01 \\
$C_{\psi'\sigma}$ [GeV$^{-2}$]
&   125.55  & $\pm$ &  5.88 \\
$C_{\psi'f}$ [GeV$^{-2}$]
&           &  $\dots$   &       \\
$\Lambda$ [GeV]
&  \multicolumn{3}{c}{3.80}  \\
\cmidrule(lr){1-4}
$\chi^2$/d.o.f.
&  \multicolumn{3}{c}{1.47}  \\
\midrule\midrule
\end{tabular}
\end{table}

For the contribution of the $\sigma(500)$ and $f_0(980)$ mesons,  described by the Flatté parametrization, preliminary fitting tests lead to the following conclusions: Only the $\psi(2S)\sigma(500)$ channel contributes.
To reduce the number of free parameters and enhance the stability of the fit, the remaining parameters in the Flatté model are fixed as follows.
The masses of $\Lambda^+_c,~\Xi_c^+,~\psi(2S)$, and $\pi^+$ are taken from the Particle Data Group (PDG)~\cite{ParticleDataGroup:2024cfk}.
The $m_{\sigma(500)}$ and $\bar{g}_{\sigma\pi}$ are derived from the relation
$\Bar{g}_{\sigma\pi}=g_{\sigma\pi\pi}^2/(8\pi m_{f}^2)$
using the values reported in Ref.~\cite{Hoferichter:2023mgy}.
The values used in this work are
$m_{\sigma(500)}=458\,\mathrm{MeV}$ and
$\bar{g}_{\sigma\pi}=2.47$.
We fix the cutoff scale at $\Lambda=3.8~\mathrm{GeV}$, at which the fit converges stably; a test of the stability of our results against the variation of $\Lambda$ is provided in Appendix~\ref{AAcutoff}.

The two resonant structures around
$\sqrt{s}=4.36$ GeV and $\sqrt{s}=4.66$ GeV
observed in the cross section data can be interpreted as being dominated by the $\psi(4S)$ and $\psi(5S)$ states~\cite{Badalian:2008dv,Segovia:2008ta,Zhao:2023hxc}, respectively.
A more comprehensive description incorporating such higher partial-wave contributions will be explored in a forthcoming study.

\subsection{The pole of $\psi(4360)$ and $\psi(4660)$}

\begin{figure}[!htbp]
\centering
\includegraphics[width=0.95\columnwidth]{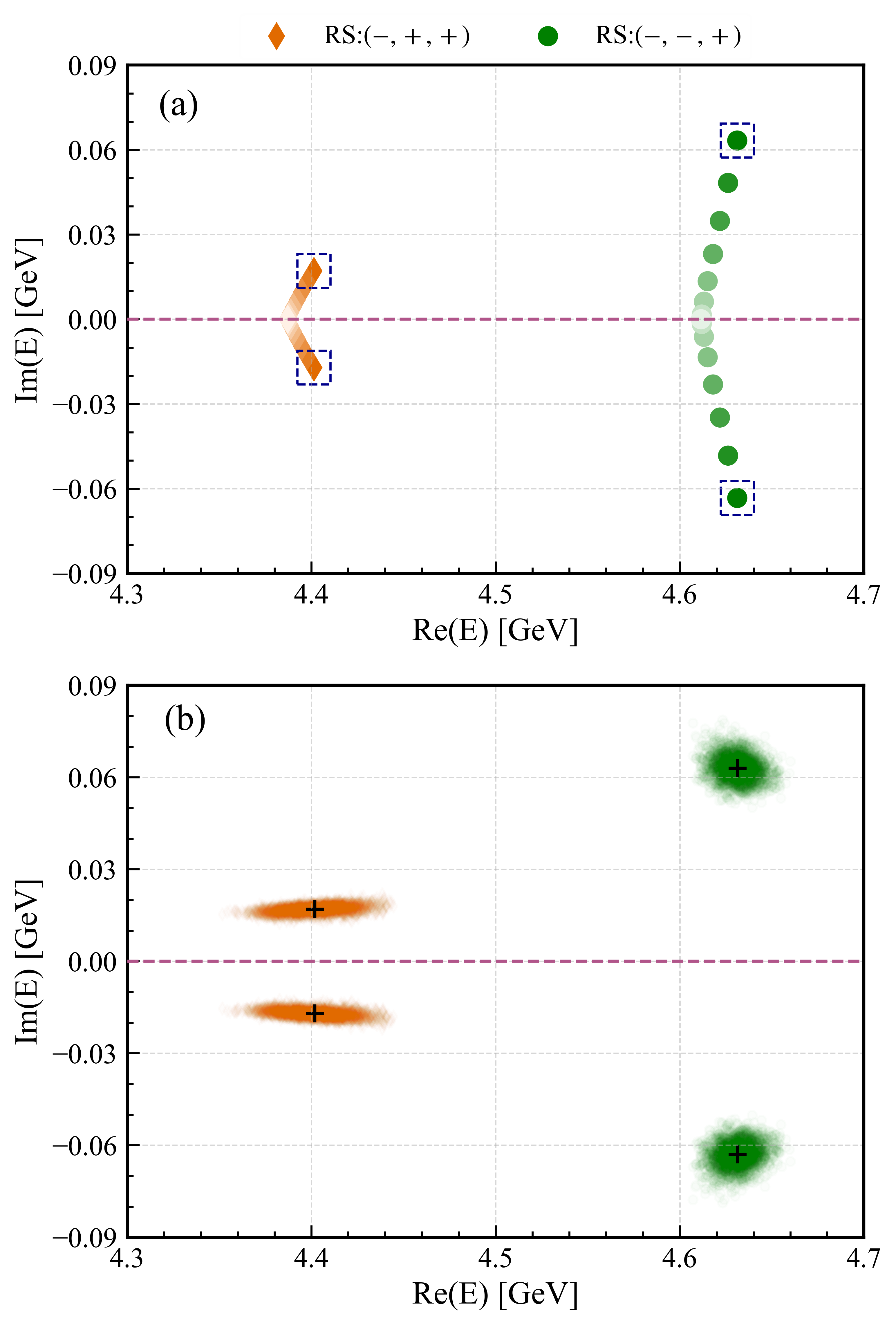}
\caption{
Panels (a): Pole trajectories of $\psi(4360)$ (orange diamonds) and $\psi(4660)$ (green circles) on different Riemann sheets (RSs), obtained using the best-fit parameters, with the coupling constants $g_{4S}$ or $g_{5S}$ gradually reduced from their fitted values to zero. The pink horizontal dashed line denotes the real axis. The blue dashed boxes indicate the initial pole positions, while the color gradient from dark to light illustrates their evolution.
Panels (b): Error regions of the pole positions on the corresponding Riemann sheets. The orange and green regions correspond to the $\psi(4360)$ and $\psi(4660)$ states, respectively. The black crosses mark the central values of the pole positions within the error intervals.
}
\label{fig:POlE}
\end{figure}

\begin{table}[!htbp]
\centering
\setlength{\tabcolsep}{10pt} 
\renewcommand{\arraystretch}{1.3}
\caption{The real parts (masses) and imaginary parts (widths) of the pole positions, together with the moduli of the dimensionless couplings obtained from our fit. The dimensionless couplings are calculated using the central values of the masses and widths.}
\label{POlEpos}
\begin{tabular}{ccc}
\midrule\midrule
        & \textbf{Pole I} & \textbf{Pole II} \\
\cmidrule(lr){1-3}
R-S&
$(-,+,+)$&$(-,-,+)$\\      
$M$ [MeV] &
$4401.93^{+15.73}_{-16.79} $ &$4631.39^{+8.27}_{-8.46}$\\     
$\Gamma$ [MeV] &
$33.93^{+2.49}_{-2.80}    $ &$125.92^{+7.88}_{-7.96}  $\\     
$|g_{\psi'\sigma}|$&
1.73 & 2.34 \\      
$|g_{\Lambda_c^+\bar{\Lambda}_c^-}|$&
1.37 & 1.83 \\      
$|g_{\Xi_c\bar{\Xi}_c}|$&
1.54 & 1.38 \\      
\midrule\midrule
\end{tabular}
\end{table}

Three types of states—bound states, virtual states, and resonances—can be identified from the pole of the $T$-matrix. Their pole positions are determined by solving the equation
\begin{align}
\textbf{det}\left[\mathbf{1}-\hat{V}_{ij}^{\rm eff}~G_{jk}(E)\right]=0.
\end{align}
Analytic continuation maps the complex $E$-plane onto a multisheeted Riemann surface consisting of $2^n$ sheets, where $n$ is the number of involved  channels in the system.
These sheets are labeled as $(\pm,\dots,\pm)$, with $\pm$ indicating the branch choice for the three-momentum in each two-body channel, i.e. $+$ ($-$) for the physical (unphysical) sheet of the corresponding channel.
The physical observables are measured on the physical Riemann sheet(R-S) $(\dots,+,\dots)$, where all the momenta are taken as $k_i=\sqrt{2\mu(E-m_{i1}-m_{i2})}$ for all the channels (see Eq.~(\ref{2PF})).
For the $\psi'\sigma$ and $\psi'f$ channels with Flatté parametrization, $k$ is given by Eq.~(\ref{flatteK}).
Further details on this Riemann sheet convention can be found in Refs.~\cite{Ye:2025ywy,Hu:2025ppi}.
Although poles could exist on all the $2^n$ Riemann sheets in principle, only those on the physical sheet $(+,\dots,+)$ and the nearby unphysical sheets
$(-,+,\dots,+)$, $(-,-,+,\dots,+)$, $\dots$, $(-,\dots,-,+,\dots,+)$, $\dots$, $(-,\dots,-)$ have significant impact on the observables.
In our analysis, the fit includes three reaction channels,
$e^+e^-\to$
$\psi'\sigma$, $\Lambda_c^+\bar{\Lambda}_c^-$, and
$\Xi_c\bar{\Xi}_c$ (ordered by their thresholds),
with the corresponding $2^3$ Riemann sheets labeled as $(\pm,\pm,\pm)$.
We have checked that the contribution of the $\psi'f$ channel is negligible and therefore is neglected.

Besides determining the pole positions, one can also extract the effective couplings $g_i$, which characterize the strength of the interaction between a given state and the $i$th channel.
$g_i$ are determined from the residues of the $T$ matrix at the given pole.
The $T$-matrix element between the $i$th and the $j$-th channels at the resonance $r$ reads as
\begin{align}
    T_{ij}(E)=\frac{g_ig_j}{E-E_r}
\end{align}
with $g_i$ the effective coupling between the given resonance $r$ and the $i$th channel. Alternatively,
one can extract the effective coupling $g_i$ from the $T$-matrix element $T_{ii}(E)$ as
\begin{align}
g_i g_i=\lim_{E\to E_r}(E-E_r)T_{ii}(E).
\end{align}
Here the pole position is written as $E_r=M_r-\iu\Gamma_r/2$ where $M_r$ and $\Gamma_r$ denote the mass and width of the state.

The pole positions of $\psi(4360)$ and $\psi(4660)$, together with their extracted effective couplings to the $i$th channel are summarized in Tab.~\ref{POlEpos}.
In our combined analysis, a pole at $4401.93\pm16.96\iu~\mathrm{MeV}$ on the $(-,+,+)$ sheet corresponds to the $\psi(4360)$ structure, while another pole at $4631.39\pm62.96\iu~\mathrm{MeV}$ on the $(-,-,+)$ sheet, located about $58~\mathrm{MeV}$ above the $\Lambda^+_c\Bar{\Lambda}^-_c$ threshold, corresponds to the $\psi(4660)$.

To distinguish the nature of the poles, we vary one of the coupling constants ($g^0_{5S}$ or $g^0_{4S}$) while keeping all other fit parameters fixed. By tracking the pole trajectories on the Riemann sheets as the couplings are gradually reduced to zero, we can characterize their origin. Specifically, if a pole moves toward the bare mass value, which is obtained from the fit, on the real axis as the coupling approaches zero, it indicates that the pole originates from a renormalized bare charmonium state. The pole trajectories obtained from our fit are illustrated in Fig.~\ref{fig:POlE}.

As illustrated in Fig.~\ref{fig:POlE}(a), when $g^0_{4S}$ or $g^0_{5S}$ is varied to zero independently, both pole $4401.93\pm16.96\iu~\mathrm{MeV}$ and $4631.39\pm62.96\iu~\mathrm{MeV}$ approach their corresponding bare masses $m_{4S}=4386.99\pm 17.10~\mathrm{MeV}$ and $m_{5S}=4611.52\pm 7.77~\mathrm{MeV}$ on the real axis, respectively.
Based on the fit, we conclude that both $\psi(4360)$ and $\psi(4660)$ originate from strong renormalization of bare vector charmonium states. These bare masses are consistent with the masses of the $4^3S_1$ and $5^3S_1$ charmonia within uncertainties, respectively,
calculated in Refs.~\cite{Badalian:2008dv,Segovia:2008ta,Zhao:2023hxc}.

\begin{figure}[!htbp]
\centering
\includegraphics[width=0.9\columnwidth]{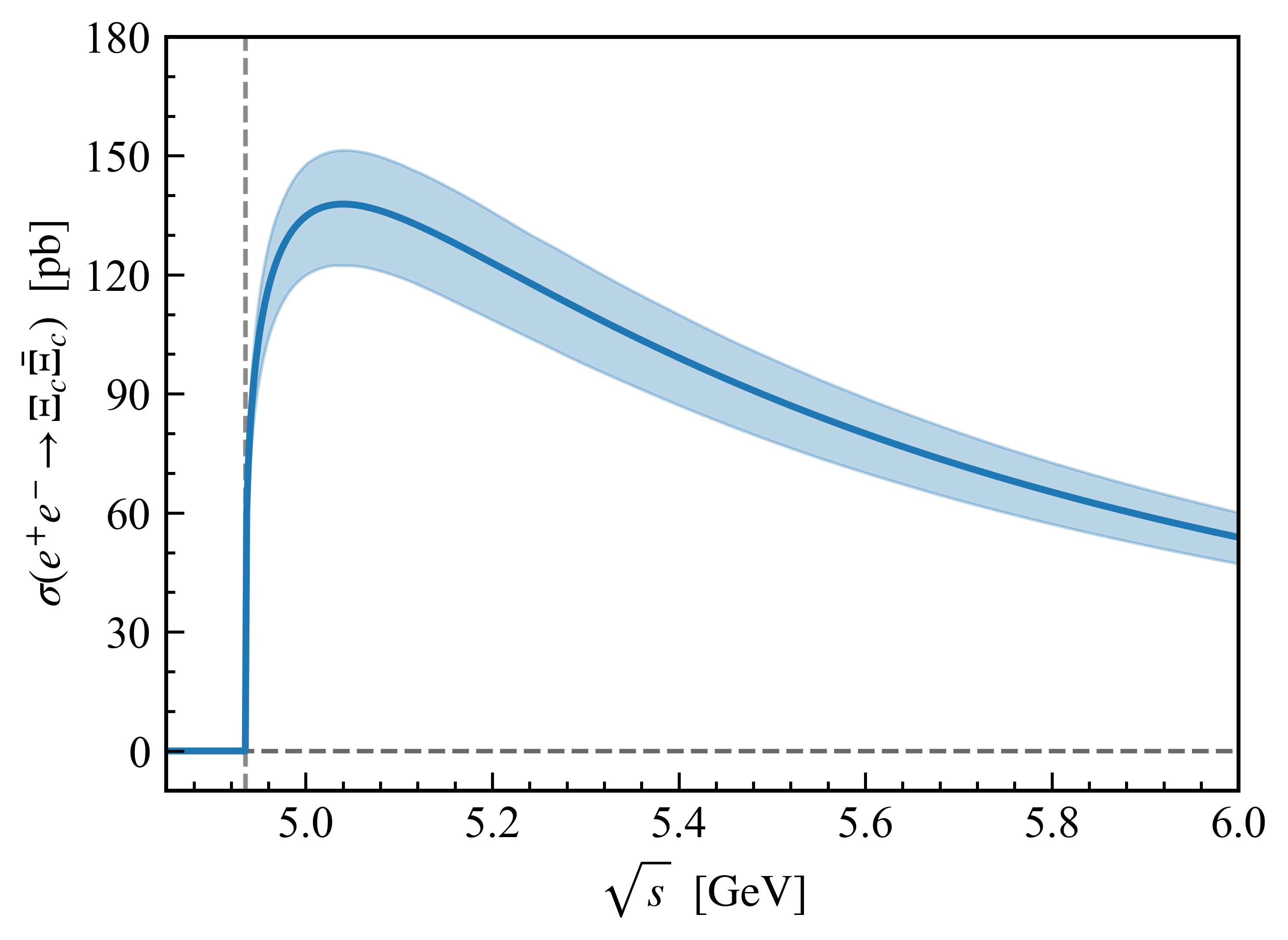}
\caption{Prediction of the $\Xi_c\bar{\Xi}_c$ cross section.}
\label{fig:Prediction}
\end{figure}

\subsection{Prediction for the $\Xi_c\bar{\Xi}_c$ cross section}

Using the fitted parameters obtained from the fit, we can predict the $e^+e^-\to\Xi_c\bar{\Xi}_c$ cross section, as shown in Fig.~\ref{fig:Prediction}. The prediction exhibits a prominent enhancement right at the $\Xi^+_c\bar{\Xi}^-_c$ threshold, stemming from the poles below threshold. Around center-of-mass energy $5.0~\mathrm{GeV}$, our prediction yields a peak cross section in the range of approximately $120-150~\mathrm{pb}$. Considering that the luminosity of BEPCII-U at a beam energy of $2.5~\mathrm{GeV}$ is about $1\times 10^{33}~\mathrm{cm}^{-2} \mathrm{s}^{-1}$, BEPCII-U will collect $430-540$ events within one hour.
Therefore, measuring the
 $e^+e^-\to\Xi_c\bar{\Xi}_c$ cross section in the near future is highly promising.

\section{SUMMARY}\label{sec:Summary}

In this work, we have addressed the long-standing puzzle of the $\psi(4360)$ and $\psi(4660)$ states, and in particular the question raised by the recent high-precision BESIII data—whether the $\psi(4660)$ exists at all. To this end, we constructed an S-wave coupled-channel framework for $e^+e^-$ annihilation
in the energy region $\sqrt{s}\in [4.0,5.0]~\mathrm{GeV}$, based on heavy-quark spin symmetry and SU(3) flavor symmetry. The formalism incorporates the open-charm baryon-antibaryon channels $\Lambda_c^+\bar{\Lambda}_c^-$, $\Xi_c^+\bar{\Xi}_c^-$, and $\Xi_c^0\bar{\Xi}_c^0$, the inelastic $\psi(2S)\pi^+\pi^-$ channel with the $\sigma(500)$ and $f_0(980)$ resonances described by Flatté parametrizations, and the bare $4S$ and $5S$ vector charmonium states, with the dynamics solved through the Lippmann-Schwinger equation. The Sommerfeld factor is explicitly included to account for the Coulomb enhancement of the charged $\Lambda_c^+\bar{\Lambda}_c^-$ pair near threshold.

A combined fit to the Belle and BESIII measurements of the $e^+e^-\to \Lambda_c^+\bar{\Lambda}_c^-$ and $e^+e^-\to \psi(2S)\pi^+\pi^-$ cross sections describes the data well ($\chi^2/\mathrm{d.o.f.}=1.47$) and yields two poles: one at $(4401.93\pm  16.96\iu)~\mathrm{MeV}$ on the $(-,+,+)$ Riemann sheet, corresponding to the $\psi(4360)$, and one at $(4631.39 \pm  62.96\iu)~\mathrm{MeV}$ on the $(-,-,+)$ sheet, corresponding to the $\psi(4660)$. The latter lies about 58~MeV above the $\Lambda_c^+\bar{\Lambda}_c^-$ threshold, and its large width naturally explains why the measured cross section appears as a nearly flat line shape rather than a sharp peak, resolving the apparent tension between the Belle and BESIII observations. By tracking the pole trajectories while the coupling to the bare charmonium states is gradually reduced to zero, we have shown that both poles continuously evolve toward the bare $4S$ and $5S$ states. This demonstrates that the $\psi(4360)$ and $\psi(4660)$ are conventional vector charmonia strongly renormalized by coupled-channel dynamics, rather than exotic or purely dynamical (molecular) states—thereby answering the title question in the affirmative and settling a controversy that persisted in earlier analyses. This conclusion is robust with respect to the choice of the regularization cutoff in the range $\Lambda=3.4\sim 4.2~\mathrm{GeV}$.

Finally, we provide predictions for the $e^+e^-\to \Xi_c\bar{\Xi}_c$ cross sections, which will be directly testable with the upcoming BEPCII upgrade scanning the energy region 4.95–5.6 GeV that covers the ground-state charmed baryon thresholds. More generally, our analysis illustrates a lesson of broad relevance to the XYZ sector: a conventional quarkonium state located near a strong S-wave open-flavor threshold can be so strongly renormalized that its experimental signature deviates drastically from naive quark-model expectations.

\section*{Acknowledgments}
We acknowledge Meng-Lin Du, Xin-Yue Hu, Pengyu Niu, and  Satoshi X. Nakamura for useful discussions. This work is partly supported by the National Natural Science Foundation of China with Grants No.~12375073 and No.~12547105.

\nocite{*}
\bibliography{ref.bib}  

\onecolumngrid
\newpage
\appendix


\section{FLATT\'{E} PARAMETRIZATION OF THE TWO-POINT FUNCTION}\label{AAfla2PF}
In this Appendix, we present the Flatté parametrization of the two-point function for the $\psi(2S)-f_0(980)$ system. The case of the $\psi(2S)-\sigma(500)$  system follows analogously.
The two-point function reads
\begin{align}
G_{\psi'f}
&=\iu\int\frac{d^4q}{(2\pi^4)}\frac{f^2(|\Vec{q}|^2)}{[q_0^2-\omega_f^2+\iu m_f(\Gamma_{\pi\pi}+\Gamma_{K\Bar{K}})][(E-q_0)^2-\omega_{\psi}^2+\iu\epsilon]}
\label{AAG1}
\end{align}
where $E$ is the center-of-mass energy, $\Gamma_{\pi\pi}$ and $\Gamma_{K\Bar{K}}$ are the relativistic partial widths of $f_0(980)$ as given in Eqs.~(\ref{PartialWidthsK})--(\ref{PartialWidthsPi}), and $\omega_j=\sqrt{|\Vec{q}|^2+m_j^2}$ with $j=\psi\,,f$.
In the nonrelativistic approximation, where $\vec{q}\to 0$, the denominator of the above expression can be rewritten as
\begin{align}
q_0^2-\omega_f^2+\iu m_f(\Gamma_{\pi\pi}+\Gamma_{K\Bar{K}})
&\approx(q_0+m_f+\frac{|\Vec{q}|^2}{2m_f})(q_0-m_f-\frac{|\Vec{q}|^2}{2m_f})+\iu m_f(\Gamma_{\pi\pi}+\Gamma_{K\Bar{K}})\notag\\
&\approx2m_f\left[q_0-m_f-\frac{|\Vec{q}|^2}{2m_f}+\frac{\iu}{2}(\Gamma_{\pi\pi}+\Gamma_{K\Bar{K}})\right]
\label{denom1}
\end{align}
Similarly,
\begin{align}
(E-q_0)^2-\omega_{\psi}^2+\iu\epsilon
\approx2m_{\psi}(E-q_0-m_{\psi}-\frac{|\Vec{q}|^2}{2m_{\psi}}+\iu\epsilon)
\label{denom2}
\end{align}
For the partial widths, we have
\begin{align}
\Gamma_{\pi\pi}=\bar{g}_{f\pi}\sqrt{\frac{q^2}{4}-m_{\pi}^2}
&=\frac{\bar{g}_{f\pi}}{2}\sqrt{\Big(q_0-\sqrt{|\vec{q}|^2+(2m_{\pi})^2}\Big)\Big(q_0+\sqrt{|\vec{q}|^2+(2m_{\pi})^2}\Big)}\notag\\
&\approx\bar{g}_{f\pi}\sqrt{m_{\pi}(q_0-2m_{\pi}-\frac{|\Vec{q}|^2}{4m_{\pi}})}
\label{Gammapipi}
\end{align}
Similarly,
\begin{align}
\Gamma_{K\Bar{K}}=\bar{g}_{f_K}\sqrt{m_K(q_0-2m_K-\frac{|\Vec{q}|^2}{4m_K})}
\label{GammaKK}
\end{align}
Plugging Eqs.~(\ref{denom1})--(\ref{denom2}) into Eq.~(\ref{AAG1}), one can obtain
\begin{align}
G_{\psi'f}
&=\frac{\iu}{4m_{\psi}m_{f}}\int\frac{d^4q}{(2\pi)^4}\frac{f^2(|\Vec{q}|^2)}{[q_0-m_f-\frac{|\Vec{q}|^2}{2m_f}+\frac{\iu}{2}(\Gamma_{\pi\pi}+\Gamma_{K\Bar{K}})][E-q_0-m_{\psi}-\frac{|\Vec{q}|^2}{2m_{\psi}}+\iu\epsilon]}\notag\\
&=\frac{1}{4m_{\psi}m_{f}}\int\frac{d^3q}{(2\pi)^3}\frac{f^2(|\Vec{q}|^2)}{E-m_{\psi}-m_f-\frac{|\Vec{q}|^2}{2m_{\psi}}-\frac{|\Vec{q}|^2}{2m_f}+\frac{\iu}{2}(\Gamma_{\pi\pi}+\Gamma_{K\Bar{K}})}
\label{AAG2}
\end{align}
Furthermore, after substituting Eqs.~(\ref{Gammapipi})--(\ref{GammaKK}) into Eq.~(\ref{AAG2}) and performing the corresponding variable redefinitions, we obtain
\begin{align}
G_{\psi'f}=\frac{2\mu_1}{4m_{\psi}m_{f}}\int\frac{d^3q}{(2\pi)^3}\frac{f^2(|\Vec{q}|^2)}{k_1^2-|\Vec{q}|^2+\iu\mu_1\left[\bar{g}_{f\pi}\sqrt{\frac{m_\pi}{2\mu_2}(k_2^2-|\Vec{q}|^2)}+\bar{g}_{f_K}\sqrt{\frac{m_K}{2\mu_3}(k_3^2-|\Vec{q}|^2)}\right]}
\label{AAG3}
\end{align}
where $\mu_1 \equiv \frac{m_f\,m_{\psi}}{m_f+m_{\psi}}$, $k_1 \equiv \sqrt{2\mu_1(E-m_{\psi}-m_f)}$, $\mu_2 \equiv \frac{2m_\pi\,m_{\psi}}{2m_\pi+m_{\psi}}$, $k_2 \equiv \sqrt{2\mu_2(E-m_{\psi}-2m_\pi)}$, $\mu_3 \equiv \frac{2m_K\,m_{\psi}}{2m_K+m_{\psi}}$, and $k_3 \equiv \sqrt{2\mu_3(E-m_{\psi}-2m_K)}$.
Ignoring the $|\Vec{q}|^2$ in partial widths, we can obtain
\begin{align}
G_{\psi'f}&\approx\frac{2\mu_1}{4m_{\psi}m_{f}}\int\frac{d^3q}{(2\pi)^3}\frac{f^2(|\Vec{q}|^2)}{k_1^2-|\Vec{q}|^2+\iu\mu_1\left[\bar{g}_{f\pi}\sqrt{\frac{m_\pi k_2^2}{2\mu_2}}+\bar{g}_{f_K}\sqrt{\frac{m_Kk_3^2}{2\mu_3}}\right]}\notag\\
&=\frac{2\mu_1}{4m_{\psi}m_{f}}\int\frac{d^3q}{(2\pi)^3}\frac{f^2(|\Vec{q}|^2)}{k^2-|\Vec{q}|^2}\label{AAG4}
\end{align}
The momentum $k$ in the above expression is evaluated by taking into account the rescattering effect through the intermediate $f_0(980)$ resonance and is given by
\begin{align}
k=\sqrt{k_1^2+\iu\mu_1\Gamma_{\rm tot}}.
\end{align}
Here, $\Gamma_{\rm tot}$ denotes the total nonrelativistic decay width, which is the sum of the partial widths into the $\pi\pi$ and $K\bar{K}$ channels and takes the following form
\begin{align}
\Gamma_{\rm tot}=\bar{g}_{f\pi}\sqrt{m_\pi(E-m_\psi-2m_\pi)}+\bar{g}_{f_K}\sqrt{m_K(E-m_\psi-2m_K)}.
\end{align}
With the Gaussian form factor $f_\Lambda(|\Vec{q}|^2)={\rm exp}(-|\Vec{q}|^2/\Lambda^2)$, Eq.~(\ref{AAG4}) can be written as
\begin{align}
G_{\psi'f}
&=\frac{1}{4m_{\psi}m_{f}}\frac{\mu}{\pi^2}\int_0^{\infty}\Big(\frac{k^2}{k^2-|\Vec{q}|^2}-1\Big){\rm exp}(-2|\Vec{q}|^2/\Lambda^2)d|\Vec{q}|\\
&=\frac{1}{4m_{\psi}m_{f}}\left[\frac{\mu k}{2\pi \iu}~\omega\Big(\frac{\sqrt{2}k}{\Lambda}\Big)-\frac{\mu\Lambda}{(2\pi)^{3/2}}\right]
\end{align}
where $\omega(z)$ is the Faddeeva function with the following form
\begin{align}
\omega(z)=\frac{2\iu z}{\pi}\int_0^\infty\frac{e^{-t^2}}{z^2-t^2}dt.
\end{align}
Besides, the Faddeeva function can be expressed in the following alternative form, which can be directly converted to the imaginary error function $\operatorname{erfi}(x)$:
\begin{align}
\omega(z)= e^{-z^2}\Big(1+\frac{2\iu}{\sqrt{\pi}}\int_0^z e^{t^2}dt\Big)=e^{-z^2}\Big[1+ \iu\text{erfi}(z)\Big]
\label{Faddeeva}
\end{align}
Applying Eq.~(\ref{Faddeeva}) to Eq.~(\ref{AAG4}), we derive the analytic expression for the Flatté parametrization of the two-point function for the $\psi(2S)-f_0(980)$ system:
\begin{align}
G_{\psi'f}=\frac{1}{4m_{\psi}m_{f}}\left\{\frac{-\mu\Lambda}{(2\pi)^{3/2}}+\frac{\mu k}{2\pi}e^{\frac{-2k^2}{{\Lambda}^2}}\left[\text{erfi}\left(\frac{\sqrt{2}k}{\Lambda}\right)-\iu\right]\right\},
\end{align}

\section{THREE-BODY PHASE SPACE INTEGRAL}\label{AA3body}

According to Eqs.~(\ref{M44})--(\ref{M55}), the squared amplitude of the $e^+e^-\to\psi(2S)\pi^+\pi^-$ channel in the Flatté parametrization can be expressed as

\begin{align}
\overline{|\mathcal{M}|}^2=\frac{1}{4}\sum_{\text{spins}}\sum_{\text{pol}}|\mathcal{M}|^2
&=\frac{e^2|\mathcal{U}|^2|f_{\rm
el}|^2}{s^2}\Big[p_+^\mu p_-^{\nu}+p_+^\nu p_-^{\mu}+g^{\mu\nu}(p_+\cdot  p_-)\Big]\left[g_{\mu\nu}-\frac{p_{1\mu}p_{1\nu}}{(p_1)^2}\right]\\
&=\frac{e^2|\mathcal{U}|^2|f_{\rm
el}|^2}{s^2}\left(s-\frac{s_2}{2}+\frac{t_1}{2}-\frac{s\,t_1}{2m_{\psi}^2}-\frac{t_1^2}{2m_{\psi}^2}+\frac{s_2\,t_1}{2m_{\psi}^2}\right),
\end{align}
where $f_{\rm el}$ denotes the $\pi\pi$ amplitude in the Flatté parametrization, as given in Eqs.~(\ref{fla500}) or (\ref{fla980}), and $\mathcal{U}$ denotes the effective production amplitude for $\psi(2S)\sigma(500)$ or $\psi(2S)f_0(980)$ channel.
We adopt the following form for the three-body phase space
\begin{align}
\Phi_3=\frac{1}{(2\pi)^5}\frac{\pi^2}{4s}\int\frac{1}{(s_2-t_1)}dt_1dt_2ds_2
\end{align}
The cross section for $e^+e^-\to\psi(2S)\pi^+\pi^-$ can be written as

\begin{align}
\sigma
&=\frac{1}{2E_{e^+}2E_{e^-}|v_{e^+}-v_{e^-}|}\frac{1}{(2\pi)^5}\frac{\pi^2}{4s}\int\frac{\overline{|\mathcal{M}|}^2}{(s_2-t_1)} dt_1dt_2ds_2\\
&=\frac{e^2}{256\pi^3 s^4}\int|\mathcal{U}|^2|f_{\rm el}|^2\frac{\left(s-\frac{s_2}{2}+\frac{t_1}{2}-\frac{s\,t_1}{2m_{\psi}^2}-\frac{t_1^2}{2m_{\psi}^2}+\frac{s_2\,t_1}{2m_{\psi}^2}\right)}{(s_2-t_1)} dt_1dt_2ds_2
\end{align}
where the difference $|v_{e^+}-v_{e^-}|$ is the relative velocity of the $e^+e^-$ colliding beams as viewed from the laboratory frame.
The physical range for each invariant is determined by~\cite{Byckling:1971vca}
\begin{align}
&t_1^{\pm}=m_{\psi}^2-\frac{1}{2}(s-s_2+m_{\psi}^2)\pm\frac{1}{2}\lambda^{\frac{1}{2}}(s,s_2,m_{\psi}^2),\\
&t_2^{\pm}=m_{\pi}^2-\frac{1}{2}(s_2-t_1)\pm\frac{1}{2s_2}(s_2-t_1)\lambda^{\frac{1}{2}}(s_2,m_{\pi}^2,m_{\pi}^2),\\
&s_2^+=(\sqrt{s}-m_{\psi})^2\quad\text{and}\quad s_2^-=4m_\pi^2,
\end{align}
where the K\"all\'en function is defined as
$\lambda(x,y,z) \equiv (x-y-z)^2-4yz$.

\section{Test of cutoff($\Lambda$) independence}\label{AAcutoff}

\begin{figure}[H]
\centering
\includegraphics[width=0.9\columnwidth]{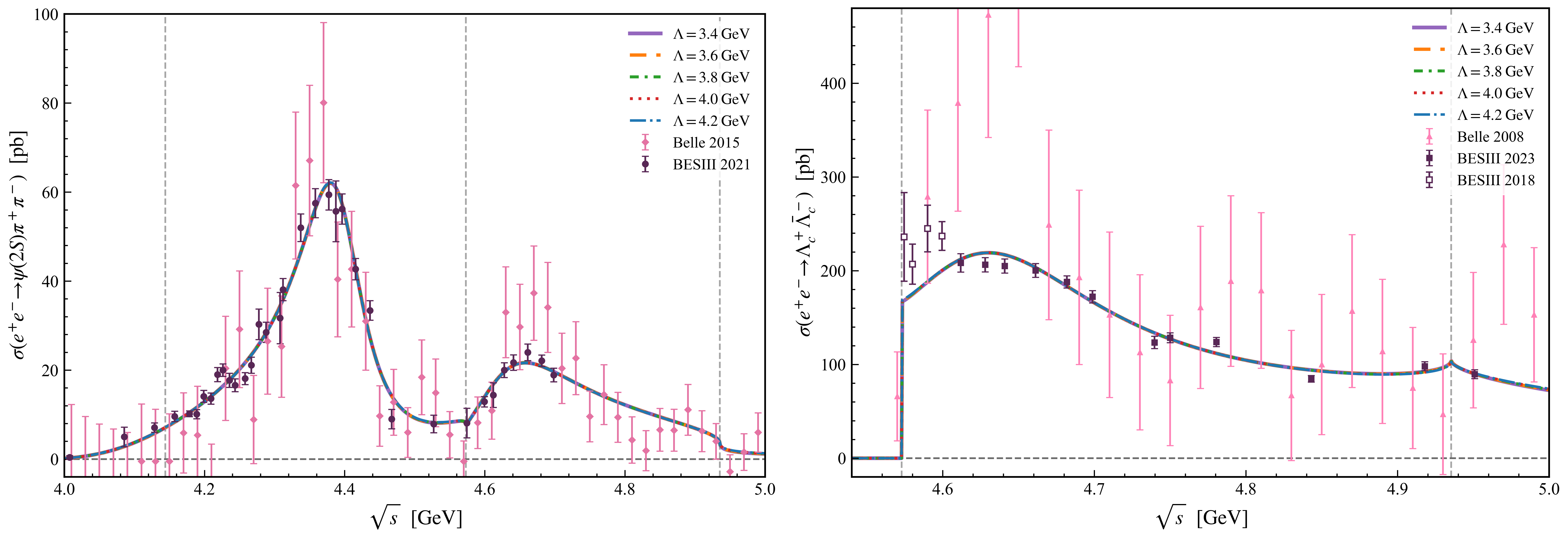}
\caption{Comparison of the fitted line shapes from the combined fit with the Belle and BESIII experimental data at various cutoffs $\Lambda$.}
\label{fig:bes_Lam}
\end{figure}

To demonstrate the independence of our results on the regularization cutoff ($\Lambda$), we repeat the combined fit with $\Lambda$ fixed at several representative values, $\Lambda=3.4$, $3.6$, $3.8$, $4.0$, and $4.2~\mathrm{GeV}$. The resulting line shapes are shown in Fig.~\ref{fig:bes_Lam}, and the corresponding fitted parameters and pole positions are collected in Tables~\ref{bespar} and~\ref{POlElam}, respectively. As can be seen, the fitted line shapes, parameters, and pole positions do not vary significantly with the value of $\Lambda$.

\begin{table}[H]
\centering
\setlength{\tabcolsep}{8pt}
\renewcommand{\arraystretch}{1.2}
\caption{The fitted parameters from the combined fit to the Belle and BESIII data at various cutoffs $\Lambda$.}
\label{bespar}
\begin{tabular}{lccccc}
\midrule\midrule
\textbf{Parameters} & $\Lambda=3.4~\mathrm{GeV}$ & $3.6~\mathrm{GeV}$ & $3.8~\mathrm{GeV}$ & $4.0~\mathrm{GeV}$ & $4.2~\mathrm{GeV}$ \\
\midrule
$g^0_{4S}$ [GeV$^{0}$]
& $-$5.26 & $-$4.98 & $-$4.73 & $-$4.52 & $-$4.32 \\
$g^0_{5S}$ [GeV$^{0}$]
& 11.05 & 10.44 & 9.89  & 9.40  & 8.96  \\
$m_{4S}$ [MeV]
& 4386.30 & 4386.71 & 4386.99 & 4387.18 & 4387.32 \\
$m_{5S}$ [MeV]
& 4610.07 & 4610.82 & 4611.52 & 4612.19 & 4612.80 \\
$f^0_{4S}$ [$10^{-1}$GeV$^{2}$]
& 0.47  & 0.47  & 0.47  & 0.47  & 0.48  \\
$f^0_{5S}$ [$10^{-1}$GeV$^{2}$]
& 0.45  & 0.45  & 0.46  & 0.46  & 0.46  \\
$f_S^0$ [$10^{-1}$GeV$^{0}$]
& 1.58  & 1.52  & 1.46  & 1.39  & 1.34  \\
$f_S^8$ [$10^{-1}$GeV$^{0}$]
& $-$5.68 & $-$5.42 & $-$5.18 & $-$4.97 & $-$4.77 \\
$C_0$ [GeV$^{-2}$]
& 94.95 & 93.11 & 91.18 & 89.25 & 87.34 \\
$C_8$ [GeV$^{-2}$]
& $-$64.30 & $-$61.38 & $-$58.74 & $-$56.34 & $-$54.14 \\
$C_{\psi'\sigma}$ [GeV$^{-2}$]
& 141.73 & 133.10 & 125.55 & 118.87 & 112.92 \\
$C_{\psi'f}$ [GeV$^{-2}$]
&  $\dots$   & $\dots$    & $\dots$    & $\dots$    & $\dots$    \\
\midrule
$\chi^2$/d.o.f.
&  1.48   &  1.48   &  1.47   &   1.47  &   1.47  \\
\midrule\midrule
\end{tabular}
\end{table}

\begin{table}[H]
\centering
\setlength{\tabcolsep}{10pt} 
\renewcommand{\arraystretch}{1.3}
\caption{Pole positions at various cutoffs $\Lambda$.}
\label{POlElam}
\begin{tabular}{ccc}
\midrule\midrule
\textbf{Cutoff($\Lambda$)} & \textbf{Pole I} & \textbf{Pole II} \\
\cmidrule(lr){1-3}
    &
$(-,+,+)$  &$(-,-,+)$  \\      
$3.4~\mathrm{GeV}$&
$4399.90\pm16.80\iu  $&$4628.92\pm61.95\iu $\\     
$3.6~\mathrm{GeV}$&
$4400.72\pm16.96\iu  $&$4630.12\pm62.65\iu $\\     
$3.8~\mathrm{GeV}$&
$4401.93\pm16.96\iu  $&$4631.39\pm62.96\iu $\\     
$4.0~\mathrm{GeV}$&
$4401.94\pm17.26\iu  $&$4632.25\pm63.84\iu $\\     
$4.2~\mathrm{GeV}$&
$4402.40\pm17.40\iu  $&$4633.20\pm64.35\iu $\\     
\midrule\midrule
\end{tabular}
\end{table}

\section{Test with a constant Breit--Wigner width for the $\sigma(500)$}\label{AAconstBW}

\begin{figure}[H]
\centering
\includegraphics[width=0.9\columnwidth]{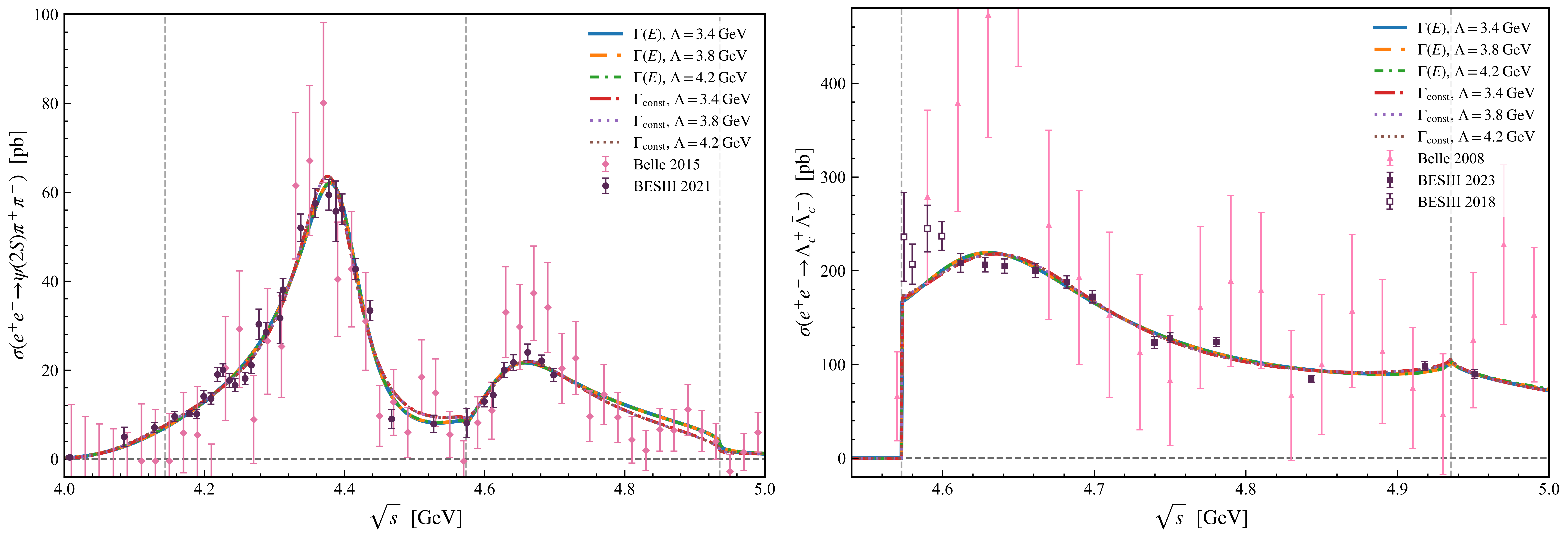}
\caption{Comparison of the fitted line shapes using the energy-dependent $(\Gamma(E))$ and constant $(\Gamma_{\rm const})$ Breit--Wigner widths for the $\sigma(500)$ at $\Lambda=3.4$, $3.8$, and $4.2$~GeV. The results show no significant qualitative difference, demonstrating that our conclusions are insensitive to the $\sigma(500)$ parametrization.}\label{fig:BWCED}
\end{figure}

To test the sensitivity of our results to the assumed line shape of the $\sigma(500)$, we repeat the combined fit by replacing its energy-dependent width by a constant Breit--Wigner width, $\Gamma^{\sigma}=558~\mathrm{MeV}$~\cite{Hoferichter:2023mgy}, for three representative values of the cutoff, $\Lambda=3.4$, $3.8$, and $4.2~\mathrm{GeV}$. The resulting line shapes are compared with those obtained in the main text in Fig.~\ref{fig:BWCED}, and the corresponding fitted parameters and pole positions are collected in Tables~\ref{ft} and~\ref{pt}, respectively. As seen from these results, the line shapes and pole positions remain qualitatively stable against this modification, confirming that our conclusions are not sensitive to the particular parametrization of the $\sigma(500)$ width.

\begin{table}[H]
\centering
\setlength{\tabcolsep}{8pt}
\renewcommand{\arraystretch}{1.2}
\caption{Comparison of the fitted parameters obtained with the energy-dependent and constant Breit--Wigner widths for the $\sigma(500)$ at $\Lambda=3.4$, $3.8$, and $4.2$~GeV.}
\label{ft}
\begin{tabular}{lcccccc}
\midrule\midrule
\textbf{Parameters} & \multicolumn{3}{c}{\textbf{Energy-dependent width}} & \multicolumn{3}{c}{\textbf{Constant width}}\\
Cutoff($\Lambda$) & $3.4~\mathrm{GeV}$ & $3.8~\mathrm{GeV}$ & $4.2~\mathrm{GeV}$ & $3.4~\mathrm{GeV}$ & $3.8~\mathrm{GeV}$ & $4.2~\mathrm{GeV}$\\
\midrule
$g^0_{4S}$ [GeV$^{0}$]
& $-$5.26 & $-$4.73 & $-$4.32
& $-$7.52 & $-$6.94 & $-$6.46 \\
$g^0_{5S}$ [GeV$^{0}$]
& 11.05 & 9.89  & 8.96
& 13.00 & 11.82 & 10.86 \\
$m_{4S}$ [MeV]
& 4386.30 & 4386.99 & 4387.32
& 4347.72 & 4348.23 & 4348.54 \\
$m_{5S}$ [MeV]
& 4610.07 & 4611.52 & 4612.80
& 4617.87 & 4619.29 & 4620.41 \\
$f^0_{4S}$ [$10^{-1}$GeV$^{2}$]
& 0.47  & 0.47  & 0.48
& 0.71  & 0.72  & 0.73 \\
$f^0_{5S}$ [$10^{-1}$GeV$^{2}$]
& 0.45  & 0.46  & 0.46
& 0.53  & 0.53  & 0.53 \\
$f_S^0$ [$10^{-1}$GeV$^{0}$]
& 1.58  & 1.46  & 1.34
& 2.15  & 2.06  & 1.98 \\
$f_S^8$ [$10^{-1}$GeV$^{0}$]
& $-$5.68 & $-$5.18 & $-$4.77
& $-$5.61 & $-$5.05 & $-$4.59 \\
$C_0$ [GeV$^{-2}$]
& 94.95 & 91.18 & 87.34
& 159.63 & 164.21 & 166.41 \\
$C_8$ [GeV$^{-2}$]
& $-$64.30 & $-$58.74 & $-$54.14
& $-$65.09 & $-$60.18 & $-$55.93 \\
$C_{\psi'\sigma}$ [GeV$^{-2}$]
& 141.73 & 125.55 & 112.92
& 169.80 & 153.85 & 141.18 \\
\midrule
$\chi^2$/d.o.f.
&  1.48   &  1.47   &   1.47
&  1.49   &  1.47   &   1.46 \\
\midrule\midrule
\end{tabular}
\end{table}

\begin{table}[H]
\centering
\setlength{\tabcolsep}{8pt} 
\renewcommand{\arraystretch}{1.3}
\caption{Comparison of the pole positions obtained with the energy-dependent and constant Breit--Wigner widths for the $\sigma(500)$ at $\Lambda=3.4$, $3.8$, and $4.2$~GeV.}
\label{pt}
\begin{tabular}{ccccc}
\midrule\midrule
             & \multicolumn{2}{c}{\textbf{Energy-dependent width}} & \multicolumn{2}{c}{\textbf{Constant width}} \\
             & \textbf{Pole I} & \textbf{Pole II} & \textbf{Pole I} & \textbf{Pole II} \\
\midrule
Cutoff~($\Lambda$)  & $(-,+,+)$  & $(-,-,+)$&
$(-,+,+)$  & $(-,-,+)$\\
$3.4~\mathrm{GeV}$&
$4399.90\pm16.80\iu  $&$4628.92\pm61.95\iu $&    
$4368.52 - 20.37\iu  $&$4640.12 - 64.93\iu $  \\
$3.8~\mathrm{GeV}$&
$4401.93\pm16.96\iu  $&$4631.39\pm62.96\iu $&    
$4370.04 - 20.49\iu  $&$4642.37 - 65.34\iu $  \\
$4.2~\mathrm{GeV}$&
$4402.40\pm17.40\iu  $&$4633.20\pm64.35\iu $&    
$4371.23 - 20.61\iu  $&$4644.19 - 65.65\iu $  \\
\midrule\midrule
\end{tabular}
\end{table}

\end{document}